\documentclass[11pt,a4paper]{article}

\usepackage[utf8]{inputenc}
\usepackage[T1]{fontenc}
\usepackage{lmodern}
\usepackage[english]{babel}
\usepackage{microtype}
\usepackage[margin=1in]{geometry}
\usepackage{amsmath,amssymb,amsthm}
\usepackage{graphicx}
\usepackage{booktabs}
\usepackage{multirow}
\usepackage{array}
\usepackage{caption}
\usepackage{subcaption}
\usepackage{xcolor}
\usepackage{tikz}
\usetikzlibrary{shapes.geometric,arrows.meta,positioning,fit,backgrounds,calc,decorations.pathreplacing}
\usepackage{algorithm}
\usepackage{algpseudocode}
\usepackage[colorlinks=true,citecolor=blue,linkcolor=blue,urlcolor=blue,linktocpage=true]{hyperref}
\usepackage{orcidlink}
\usepackage{doi}
\usepackage[capitalize]{cleveref}
\usepackage{enumitem}
\usepackage{titlesec}
\usepackage{authblk}
\usepackage{footmisc}
\usepackage[font=small,labelfont=bf]{caption}
\usepackage{siunitx}
\title{\Large\bfseries
Latency-Constrained Hardware-Aware Quantum Error Correction Co-Design with Adaptive Confidence-Gated Neural Decoding for the Rotated Surface Code}

\author[1]{Sumit Chongder\orcidlink{0009-0005-9866-8483}}
\affil[1]{Department of Physics, Quantum Information and Computation, Indian Institute of Technology Jodhpur, Jodhpur, Rajasthan 342037, India}
\affil[ ]{\textit{Email:} \texttt{sumitchongder960@gmail.com}}

\date{}

\begin{document}

\maketitle
\begin{abstract}
\noindent
Real-time decoding is a major bottleneck in scaling quantum error correction (QEC) from present-day noisy intermediate-scale quantum (NISQ) devices to fault-tolerant quantum computing. Although neural decoders can approach the accuracy of minimum-weight perfect matching (MWPM) for small code distances, average decoding accuracy alone is insufficient for practical deployment because rare, low-confidence syndromes disproportionately contribute to logical failures. We present an adaptive confidence-gated decoding framework for the rotated surface code that treats decoding as a two-stage inference problem. A lightweight feed-forward neural network performs fast-path decoding for the majority of syndrome measurements, while only low-confidence predictions are escalated to an MWPM refinement stage. This adaptive decoder constitutes the first implemented component of a broader hardware-aware QEC co-design architecture aimed at jointly optimizing decoding performance and deployment efficiency. We benchmark the framework on rotated surface codes with distances $d\in\{3,5,7,9,11\}$ under circuit-level depolarising noise using the Stim stabiliser simulator. The evaluation systematically characterises logical accuracy, logical error-rate scaling, confidence-controlled accuracy--latency trade-offs, decoding throughput, per-shot latency, and decoding-graph resource scaling across code distances, physical error rates, and inference batch sizes. Routing only $3.3\%$--$6.2\%$ of syndromes to the refinement stage improves logical accuracy from $99.21\%$ for the neural-only baseline to $99.81\%$ at a confidence threshold of $0.95$, while incurring only a bounded increase in average decoding cost. We further observe neural-decoder throughput saturating near $4.6\times10^{5}$ samples\,s$^{-1}$ at batch size $512$ on commodity CPU hardware, indicating that the neural fast path is not the dominant throughput bottleneck beyond code distance $d=7$. To support reproducible research, we release the complete benchmarking pipeline, trained models, raw benchmark data, and source code. Finally, we explicitly distinguish the experimentally validated contributions of this work from the broader hardware-aware co-design roadmap, including hardware-constrained code discovery, GPU-accelerated inference, and multi-noise optimisation, which we identify as directions for future research.
\end{abstract}


\newpage
\tableofcontents

\section{Introduction}
\label{sec:intro}

Quantum error correction (QEC) is the only known route to running quantum algorithms whose circuit depth and qubit count exceed what any physical qubit can tolerate without active correction~\cite{Shor1995,Steane1996,Kitaev2003}. Among the family of topological codes, the surface code has become the de facto standard for near-term fault-tolerant architectures because of its high error threshold, planar nearest-neighbour connectivity, and compatibility with superconducting and neutral-atom hardware~\cite{Kitaev2003,Dennis2002,Fowler2012}. Two independent experimental milestones from Google Quantum AI have now demonstrated that increasing the surface-code distance suppresses the logical error rate exponentially in hardware, first for repetition-code memories~\cite{GoogleQuantumAI2021} and subsequently for a full two-dimensional surface-code logical qubit below the break-even threshold~\cite{GoogleQuantumAI2023}. These results shift the central engineering bottleneck of fault tolerance away from the physical qubit and toward the classical control stack that must process syndrome measurements and issue corrections within the coherence and duty-cycle constraints of the underlying hardware~\cite{Terhal2015,Campbell2017}.

Decoding a syndrome, that is, inferring the most likely physical error consistent with a set of stabiliser measurement outcomes, is an inference problem whose classical hardness scales with code distance. For the surface code under a Pauli noise model, this problem maps onto minimum-weight perfect matching (MWPM) on a decoding graph~\cite{Dennis2002,Fowler2012,Higgott2022}, or, more generally, onto belief propagation with ordered statistics decoding (BP+OSD) for quantum low-density parity-check (QLDPC) codes~\cite{Panteleev2021,Roffe2020}. Union-find decoders~\cite{Delfosse2021} and sparse-blossom implementations of MWPM~\cite{HiggottGidney2023} have pushed classical decoding latency down substantially, and window-based and hierarchical decoding schemes have been proposed to keep decoding latency bounded as code distance grows~\cite{Skoric2023,Delfosse2020,Tan2023}. In parallel, a distinct line of work has explored machine-learned decoders, beginning with feed-forward neural networks for small surface codes~\cite{Torlai2017,Varsamopoulos2017}, recurrent and convolutional architectures for streaming decoding~\cite{Baireuther2018,Chamberland2018}, and, most recently, transformer-based decoders that match or exceed correlated-noise MWPM accuracy on real device data~\cite{Bausch2024}. Neural decoders are attractive because their inference cost is dominated by a small number of fixed matrix multiplications rather than a combinatorial graph search, which makes them natural candidates for constant-latency, hardware-accelerated deployment~\cite{Overwater2022,Meinerz2022}.

However, an accuracy-only comparison between neural decoders and MWPM obscures an operationally important asymmetry. Averaged accuracy is dominated by the overwhelming majority of syndromes that are easy to decode, whereas the logical error rate of a fault-tolerant computation is dominated by the rare, hard syndromes on which any single fixed decoder is most likely to make a correlated or catastrophic error~\cite{Chamberland2018,Baireuther2018}. A decoder that is $99.5\%$ accurate on average may still be an unacceptable choice if its failures are concentrated on syndromes that a slower, more careful decoder would have resolved correctly. This observation motivates treating decoding not as a single-model classification problem, but as a \emph{control} problem: a fast, cheap decoder should handle the bulk of traffic, while a bounded, confidence-gated fraction of hard cases should be escalated to a more expensive but more reliable refinement stage~\cite{Overwater2022}. This idea is closely related to cascade and mixture-of-experts architectures in classical machine learning~\cite{Viola2001,Shazeer2017}, but it has not, to our knowledge, been characterised systematically for surface-code decoding across the joint axes of code distance, physical noise strength, escalation threshold, and inference batch size that jointly determine whether such a scheme is deployable on real control hardware.

At the same time, the decoder is only one half of a genuinely hardware-aware fault-tolerant stack. The choice of code, its embedding onto a physical qubit connectivity graph, and the classical control schedule that reads out syndromes and applies corrections are coupled design variables: a code that is asymptotically optimal in the abstract may be undeployable if its stabiliser weight, ancilla overhead, or measurement schedule exceeds what a given hardware platform supports within its coherence time~\cite{Bravyi2024,Panteleev2021}. Recent work has begun to explore reinforcement-learning and differentiable search methods for discovering codes and encodings under explicit hardware constraints~\cite{RLCodeDiscovery2023,Chamberland2018}, and vendor-level tooling such as NVIDIA's CUDA-Q QEC library has begun to expose decoding and simulation primitives as GPU-accelerated, composable building blocks for exactly this kind of co-design study~\cite{NvidiaCudaQX2024}. A unified framework that jointly reasons about code structure, decoder policy, and hardware latency budget, rather than optimising each in isolation, is therefore a natural and, we argue, necessary direction for the field.

\subsection{Contributions and scope}
\label{sec:contributions}

This paper makes the following contributions, and we state them precisely in order to be explicit about what is empirically demonstrated versus what is proposed as a design target for future work, a distinction we consider essential to responsible reporting in a fast-moving subfield.

\begin{enumerate}[leftmargin=1.5em]
\item \textbf{An adaptive, confidence-gated decoding architecture.} We design and implement a two-tier decoder for the rotated surface code in which a compact feed-forward neural network produces a per-syndrome correction together with a calibrated confidence score; syndromes whose confidence falls below a tunable threshold $\tau$ are escalated to a MWPM refinement stage implemented with PyMatching~\cite{Higgott2022}. We describe the training procedure, the confidence calibration used to make the escalation decision meaningful, and the resulting escalation-controlled accuracy-cost trade-off curve (\Cref{sec:methods:decoder,sec:results:routing}).

\item \textbf{A systematic, multi-axis empirical characterisation.} We benchmark the framework across code distance $d\in\{3,5,7,9,11\}$, six values of circuit-level depolarising error probability spanning $p\in[10^{-4},5\times10^{-3}]$, five escalation thresholds $\tau\in\{0.60,0.70,0.80,0.90,0.95\}$, and inference batch sizes spanning $1$ to $512$, reporting accuracy, logical error rate, per-shot latency, throughput, and decoder confidence distributions at every operating point (\Cref{sec:results}). To our knowledge this is among the more complete joint characterisations of a confidence-routed neural surface-code decoder reported to date, and we release the complete raw benchmark tables in the supplementary material (\Cref{app:tables}).

\item \textbf{An explicit resource-scaling analysis of the decoding graph.} We quantify how the number of stabiliser detectors, circuit operations, and estimated decoding-graph memory grow with code distance for the rotated surface code family used throughout this work, connecting the decoder-level latency results to the underlying combinatorial structure that a hardware-aware code-discovery module would need to reason about (\Cref{sec:results:resources}).

\item \textbf{A hardware-aware co-design roadmap.} We situate the adaptive decoder within a larger, explicitly-scoped framework that couples code generation, circuit-level noise simulation, and hardware-constrained ranking (\Cref{sec:methods:framework}), and we specify, as a concrete and falsifiable research programme rather than a completed result, the reinforcement-learning-based code-discovery and GPU-accelerated deployment components that would complete the closed loop described in the framework diagram of \Cref{fig:framework}. We are explicit that the code-discovery and multi-family noise-model components of this roadmap are architectural proposals evaluated only at the design level in the present paper, and we identify this as the primary direction for follow-up work in \Cref{sec:limitations}.
\end{enumerate}

We emphasise at the outset that every quantitative result reported in \Cref{sec:results} was obtained under a single code family (the rotated surface code) and a single noise family (independent circuit-level depolarising noise, as generated by Stim's standard noise model~\cite{Gidney2021}). We do not claim, and the data presented here do not support, generalisation to correlated, coherent, or leakage-dominated noise, nor to QLDPC or biased-noise code families; we return to this scope boundary explicitly in \Cref{sec:limitations}.

\subsection{Paper organisation}

\Cref{sec:related} reviews decoding algorithms, neural decoders, and hardware-aware co-design literature. \Cref{sec:methods} describes the rotated surface code construction, the circuit-level noise model, the adaptive decoder architecture and training procedure, and the full hardware-aware co-design framework within which the decoder is situated. \Cref{sec:experiments} describes the experimental protocol. \Cref{sec:results} reports our empirical findings across all benchmark axes. \Cref{sec:discussion} discusses implications, and \Cref{sec:limitations} states limitations and threats to validity explicitly. \Cref{sec:conclusion} concludes. \Cref{app:algorithms,app:tables,app:hyperparameters,app:reproducibility} provide algorithmic pseudocode, complete raw result tables, hyperparameter settings, and reproducibility information, respectively.

\section{Related work}
\label{sec:related}

\subsection{Classical decoding algorithms}

For the surface code under an independent Pauli noise model, decoding reduces to finding a minimum-weight perfect matching on a graph whose vertices are triggered stabiliser detectors and whose edge weights encode the log-likelihood of the corresponding error chain~\cite{Dennis2002,Fowler2012}. Blossom-algorithm implementations of MWPM decoding have long been the accuracy baseline for surface-code simulation studies~\cite{Fowler2012}, and the recent \emph{sparse blossom} formulation implemented in PyMatching~2 reduced MWPM decoding time by roughly two orders of magnitude relative to earlier implementations while preserving optimality guarantees under the matching-graph model~\cite{Higgott2022,HiggottGidney2023}. Union-find decoders trade a small amount of accuracy for near-linear-time decoding and have been demonstrated as a practical alternative when strict real-time budgets are required~\cite{Delfosse2021}. For general QLDPC codes, belief propagation with ordered statistics post-processing (BP+OSD) remains the standard decoding approach~\cite{Panteleev2021,Roffe2020}, and recent bivariate-bicycle QLDPC constructions have demonstrated substantially reduced physical-to-logical qubit overhead relative to the surface code at comparable distance, at the cost of longer-range connectivity requirements~\cite{Bravyi2024}. As code distance and, correspondingly, decoding-graph size grow, window-based and hierarchical decomposition schemes have been proposed to keep the latency of any of these decoders bounded within a fixed syndrome-extraction cycle time~\cite{Skoric2023,Delfosse2020,Tan2023,Ravi2023}.

\subsection{Neural and machine-learned decoders}

Torlai and Melko first demonstrated that a feed-forward neural network could learn to decode small topological codes directly from labelled syndrome data without an explicit graph model of the noise~\cite{Torlai2017}, and Varsamopoulos \emph{et al.} extended this to systematic accuracy comparisons against MWPM on the surface code~\cite{Varsamopoulos2017}. Baireuther \emph{et al.} and Chamberland and Ronagh subsequently introduced recurrent and convolutional architectures capable of streaming decoding across repeated syndrome-extraction rounds and of handling measurement errors and correlated noise more directly than a single-shot feed-forward model~\cite{Baireuther2018,Chamberland2018}.Meinerz \emph{et al.} introduced a scalable neural decoder for topological surface codes that combines a neural-network-based local correction stage with a separate global decoding stage, enabling a decoder trained on small local patches to scale efficiently across larger code distances without retraining the entire model~\cite{Meinerz2022}. In a complementary direction, Lange \emph{et al.} proposed a graph neural network (GNN) decoder for surface codes, demonstrating generalisation across code distances and decoding performance competitive with, and in some circuit-level noise settings exceeding, minimum-weight perfect matching (MWPM)~\cite{Lange2025}. Overwater \emph{et al.} performed an explicit hardware cost-performance exploration of neural-network decoders, quantifying the accuracy that different network sizes could sustain under fixed latency and area budgets on realistic control electronics, a study whose design goals are close in spirit to the present work~\cite{Overwater2022}. Most recently, Bausch \emph{et al.} introduced AlphaQubit, a transformer-based decoder trained partly on real superconducting-processor data that matches or exceeds correlated-noise MWPM decoding accuracy at distances up to $d=11$~\cite{Bausch2024}, demonstrating that learned decoders are now competitive with classical baselines on real experimental syndromes rather than only in simulation.

None of the works above, to our knowledge, systematically reports a confidence-gated two-tier decoder's accuracy-versus-escalation-fraction trade-off jointly with its throughput and latency scaling across a wide sweep of code distances, noise strengths, and batch sizes, which is the empirical gap the present paper addresses. Overwater \emph{et al.}'s hardware-cost framing~\cite{Overwater2022} and the general cascade/mixture-of-experts literature in classical machine learning~\cite{Viola2001,Shazeer2017} are the closest conceptual antecedents to our escalation mechanism, and we build directly on their framing while specialising it to the surface-code decoding setting and reporting the full multi-axis benchmark that we believe is necessary to assess deployability.

\subsection{Hardware-aware and co-design approaches to fault tolerance}

A growing literature argues that code selection, decoder design, and control-hardware constraints should not be optimised independently. Reinforcement-learning and differentiable search methods have been proposed for discovering error-correcting codes and encodings under explicit connectivity and gate-set constraints~\cite{RLCodeDiscovery2023}, and hierarchical decoding architectures have been motivated explicitly by the observation that a single monolithic decoder cannot simultaneously satisfy the accuracy and latency requirements of a large fault-tolerant device~\cite{Delfosse2020,Ravi2023}. On the tooling side, the Stim stabiliser-circuit simulator~\cite{Gidney2021} and the CUDA-Q QEC library~\cite{NvidiaCudaQX2024} have made it substantially more practical to generate large labelled syndrome datasets and to prototype GPU-accelerated decoding and simulation pipelines, respectively; our benchmarking pipeline is built directly on Stim for data generation and on PyMatching for the refinement-stage decoder~\cite{Gidney2021,Higgott2022}. We view the present paper's contribution as an empirically grounded first module, the adaptive decoder and its full benchmark characterisation, within the larger hardware-aware co-design programme that this literature motivates, and we describe explicitly in \Cref{sec:methods:framework} and \Cref{sec:limitations} which components of that programme remain to be built and evaluated.

\section{Methods}
\label{sec:methods}

\subsection{Hardware-aware co-design framework: system overview}
\label{sec:methods:framework}

\begin{figure}[htbp]
\centering
\resizebox{0.98\textwidth}{!}{%
\begin{tikzpicture}[
    node distance=1.2cm and 1.5cm,
    block/.style={draw=green!50!black,fill=green!10,rectangle,minimum width=2.8cm,minimum height=0.9cm,align=center,line width=0.8pt},
    grey/.style={draw=gray!70,fill=gray!15,rectangle,minimum width=2.8cm,minimum height=0.9cm,align=center,line width=0.8pt,dashed},
    feedback/.style={dashed, ->, line width=0.8pt},
    flow/.style={->, line width=0.8pt}
]

    \node[block] (nm) {Noise Model};
    \node[block, below=of nm] (hwc) {H/W Constraints};
    \node[grey, below=of hwc] (scg) {Surface Code Generator};
    \node[block, below=of scg] (scs) {Stim Circuit Synthesis};
    
    \node[block, left=of scs] (ss) {Stim Simulation};
    \node[block, above=of ss] (and) {Adaptive Neural Decoder};
    \node[block, above=of and] (conf) {Confidence};
    \node[block, above=of conf] (la) {Latency-aware};
    \node[block, above=of la] (gr) {Graph Refinement};
    
    \node[block, left=of gr] (lc) {Logical Correction};
    \node[block, below=of lc] (lm) {Logical Metrics};
    
    \node[grey, dashed, left=of lc] (hco) {Hardware-constrained\\optimization};

    \draw[flow] (nm) -- (hwc);
    \draw[flow] (hwc) -- (scg);
    \draw[flow] (scg) -- (scs);
    \draw[flow] (scs) -- (ss);
    \draw[flow] (ss) -- (and);
    \draw[flow] (and) -- (conf);
    \draw[flow] (conf) -- (la);
    \draw[flow] (la) -- (gr);
    \draw[flow] (gr) -- (lc);
    \draw[flow] (lc) -- (lm);

    \draw[feedback] (hco.north) |- ++(0, 0.8) -| (nm.north);
    
    \draw[feedback] (hco.south) |- ++(0, -5.8) |- (and.west);

\end{tikzpicture}
}
\caption{Hardware-aware quantum error correction co-design framework. Green solid boxes denote components implemented and empirically evaluated in this paper (noise model, circuit synthesis via Stim, adaptive confidence-gated decoder, and logical-metric evaluation). Grey dashed boxes denote the hardware-constrained code-generation and closed-loop latency-constrained optimisation components that are specified architecturally as the target of ongoing and future work (\Cref{sec:limitations}) and are not evaluated empirically here.}
\label{fig:framework}
\end{figure}
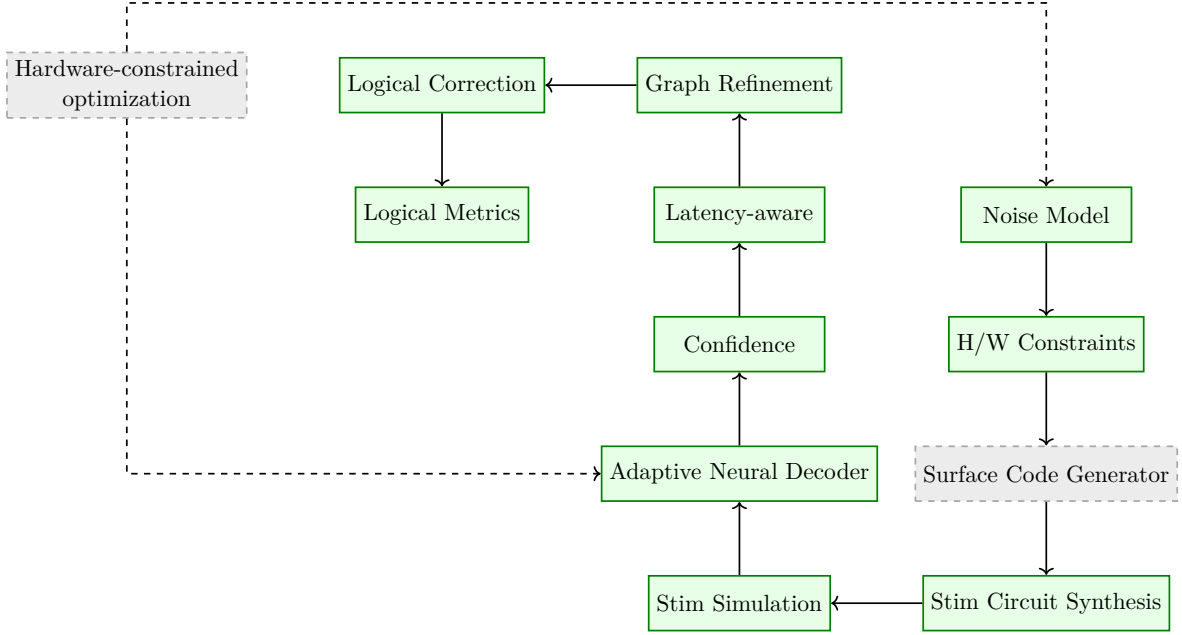

\begin{figure}[htbp]
\centering
\begin{tikzpicture}[
  node distance=7mm and 16mm,
  box/.style={draw, rounded corners=2pt, minimum width=52mm, minimum height=9mm,
              align=center, font=\small, fill=blue!6, thick},
  demo/.style={box, fill=green!8, draw=green!45!black},
  proposed/.style={box, fill=gray!12, draw=gray!55, dashed},
  arr/.style={-{Latex[length=2.2mm]}, thick}
]
\node[demo] (hw) {Hardware Constraints\\\footnotesize Connectivity $\cdot$ Gate Set $\cdot$ Cycle Time};
\node[demo, below=of hw] (noise) {Noise Model\\\footnotesize Circuit-level depolarising, $p\in[10^{-4},5\times10^{-3}]$};
\node[proposed, below=of noise] (code) {Code Generator\\\footnotesize Rotated Surface Code, $d\in\{3,5,7,9,11\}$};
\node[demo, below=of code] (circ) {Circuit Compilation\\\footnotesize Stim-compatible detector circuits};
\node[demo, below=of circ] (synd) {Syndrome Simulation\\\footnotesize $5\times10^4$ shots per operating point};
\node[demo, below=of synd] (dec) {Adaptive Decoder\\\footnotesize Neural fast path + MWPM refinement};
\node[demo, below=of dec] (eval) {Evaluation\\\footnotesize Logical error $\cdot$ Runtime $\cdot$ Throughput};

\node[proposed, right=28mm of circ] (mo) {Multi-objective\\Optimiser};
\node[proposed, below=of mo] (hp) {Updated\\Hyperparameters};

\draw[arr] (hw) -- (noise);
\draw[arr] (noise) -- (code);
\draw[arr] (code) -- (circ);
\draw[arr] (circ) -- (synd);
\draw[arr] (synd) -- (dec);
\draw[arr] (dec) -- (eval);
\draw[arr] (dec.west) -- ++(-1,0) -- ++(0,+3.4) -- (circ.west);
\draw[arr] (mo) -- (hp);
\draw[arr, dashed] (hp.south) |- (dec.east);
\draw[arr, dashed] (synd.east) -- ++(1.5,0) -- ++(0,+1.7) -- (mo.west);
\end{tikzpicture}
\caption{Hardware-aware co-design optimisation pipeline. The vertical chain (noise model, circuit compilation, syndrome simulation, adaptive decoding, evaluation) is the pipeline implemented and benchmarked in this paper. The hardware-constrained code generator and the multi-objective optimisation loop that would close the search over code candidates (right, dashed) are specified as the target architecture for the next phase of this research programme and are not exercised in the current experiments.}
\label{fig:pipeline}
\end{figure}
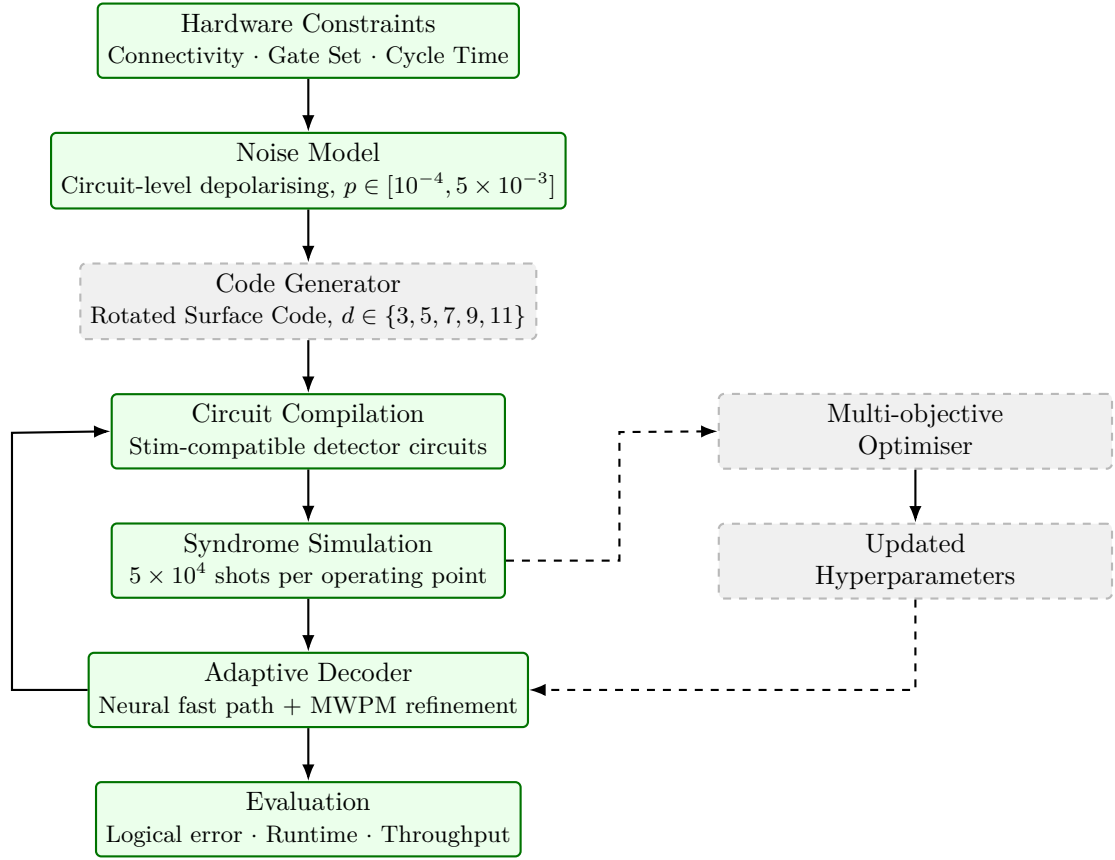

\Cref{fig:framework} shows the overall framework within which the present study is situated. The framework couples four functional stages: a \emph{code generator} that instantiates a stabiliser code template subject to hardware connectivity, gate-set, and cycle-time constraints; a \emph{noise model} that specifies the physical error channel(s) under which the code will be evaluated; a \emph{circuit-level syndrome simulator}, for which we use Stim~\cite{Gidney2021}, that compiles the code and noise model into a sampled detector-event stream; and an \emph{adaptive decoder} that consumes this stream and produces logical corrections together with per-shot latency and confidence telemetry. A hardware-aware ranking stage aggregates logical error rate, decoding latency, and resource cost into a scalar or Pareto-front objective that could in principle drive an outer code-and-hyperparameter search loop, shown as the latency-constrained optimiser in \Cref{fig:pipeline}.

In the present paper, the code generator is fixed to the rotated surface code family (\Cref{sec:methods:code}), the noise model is fixed to independent circuit-level depolarising noise (\Cref{sec:methods:noise}), and our empirical contribution is a complete characterisation of the adaptive decoder stage and the resource-scaling properties of the resulting decoding graphs (\Cref{sec:methods:decoder,sec:results:resources}). The code-generator search loop and the multi-noise-family ranking stage, both shown greyed out relative to the demonstrated components in \Cref{fig:pipeline}, are specified at the architectural level as the concrete next stage of this research programme and are discussed as such in \Cref{sec:limitations}.

\Cref{fig:pipeline} shows the same system redrawn as a data and control flow, making explicit the multi-objective ranking function
\begin{equation}
\mathcal{L} = w_1 P_L + w_2 T + w_3 M + w_4 C,
\label{eq:objective}
\end{equation}
where $P_L$ is logical error rate, $T$ is decoding runtime, $M$ is decoder memory footprint, and $C$ is a hardware deployment cost proxy (e.g.\ ancilla or connectivity overhead), and $w_1,\dots,w_4 \ge 0$ are user-specified weights. \Cref{eq:objective} defines the intended optimisation target of the full framework; in the present paper we report $P_L$, $T$, and a memory proxy for $M$ directly as multi-dimensional benchmark results (\Cref{sec:results}) rather than collapsing them into a single scalar, since we consider the resulting Pareto structure more informative than any particular choice of weights, and because, as we discuss in \Cref{sec:results:pareto}, our current benchmark grid does not yet span a resource-cost axis $C$ wide enough to make the scalarised objective meaningful.

\subsection{Code family and circuit construction}
\label{sec:methods:code}

We use the rotated surface code, a distance-$d$ planar stabiliser code encoding one logical qubit in $n=d^2$ data qubits with $n-1$ weight-four (bulk) and weight-two (boundary) stabiliser generators, arranged so that $X$-type and $Z$-type stabilisers are measured on alternating checkerboard plaquettes~\cite{Fowler2012,Tomita2014}. We evaluate $d\in\{3,5,7,9,11\}$, corresponding to $9,25,49,81,121$ data qubits respectively. Syndrome-extraction circuits are compiled to native two-qubit gates using Stim's standard rotated-surface-code circuit generator~\cite{Gidney2021}, with one round of stabiliser measurement per code cycle and $d$ rounds of syndrome extraction per logical memory experiment, following the standard memory-experiment convention used to define the logical error rate per code cycle~\cite{Fowler2012,GoogleQuantumAI2021}.

\subsection{Noise model}
\label{sec:methods:noise}

Throughout this paper we use Stim's standard circuit-level depolarising noise model~\cite{Gidney2021}: each single-qubit gate location is followed by a depolarising channel of strength $p$, each two-qubit gate location is followed by a two-qubit depolarising channel of strength $p$, and each measurement is subject to a classical readout-flip probability $p$. We sweep $p\in\{10^{-4},2\times10^{-4},5\times10^{-4},10^{-3},2\times10^{-3},5\times10^{-3}\}$, spanning roughly one and a half orders of magnitude below and around the surface code's known circuit-level threshold of approximately $p_{\mathrm{th}}\approx 0.5$--$1\%$ under this noise model~\cite{Fowler2012,Wang2011}. We do not, in the present study, evaluate coherent, leakage, or correlated (crosstalk-like) noise channels; this is a deliberate scope restriction that we discuss in \Cref{sec:limitations}, since it is these non-Pauli and correlated noise regimes in which neural decoders have been argued to offer the largest advantage over graph-based decoders~\cite{Baireuther2018,Bausch2024}, and our present results should accordingly be read as characterising decoder deployability under the simpler, better-understood depolarising setting.

\subsection{Adaptive confidence-gated decoder}
\label{sec:methods:decoder}

\begin{figure}[htbp]
\centering
\begin{tikzpicture}[
  node distance=8mm and 18mm,
  box/.style={draw, rounded corners=2pt, minimum width=44mm, minimum height=9mm,
              align=center, font=\small, fill=blue!6, thick},
  dec/.style={draw, diamond, aspect=2.4, minimum width=44mm, minimum height=16mm,
              align=center, font=\small, fill=orange!12, thick, inner sep=1pt},
  arr/.style={-{Latex[length=2.2mm]}, thick}
]
\node[box] (synd) {Measured Syndrome $\mathbf{s}$};
\node[box, below=of synd] (nn) {Feed-Forward Neural Decoder $f_\theta$};
\node[box, below=of nn] (prob) {Prediction $\hat{y}$ \& Confidence $c=\max_k p_k$};
\node[dec, below=of prob] (chk) {$c \ge \tau$?};
\node[box, below left=14mm and -6mm of chk] (fast) {Fast-Path Correction\\\footnotesize (accept $\hat y$)};
\node[box, below right=14mm and -6mm of chk] (mwpm) {MWPM Refinement\\\footnotesize (PyMatching, exact)};
\node[box, below=26mm of chk] (apply) {Apply Recovery Operator};
\node[box, below=of apply] (next) {Next Syndrome-Extraction Round};

\draw[arr] (synd) -- (nn);
\draw[arr] (nn) -- (prob);
\draw[arr] (prob) -- (chk);
\draw[arr] (chk.west) -- node[above,sloped,font=\scriptsize]{yes} (fast.north);
\draw[arr] (chk.east) -- node[above,sloped,font=\scriptsize]{no} (mwpm.north);
\draw[arr] (fast.south) |- (apply.west);
\draw[arr] (mwpm.south) |- (apply.east);
\draw[arr] (apply) -- (next);
\end{tikzpicture}
\caption{Decision flow of the adaptive, confidence-gated decoder. Every measured syndrome is first passed through the fast feed-forward decoder; syndromes with confidence at or above the operating threshold $\tau$ are corrected on the fast path, while low-confidence syndromes are escalated to exact minimum-weight perfect matching refinement before recovery is applied.}
\label{fig:decoderflow}
\end{figure}
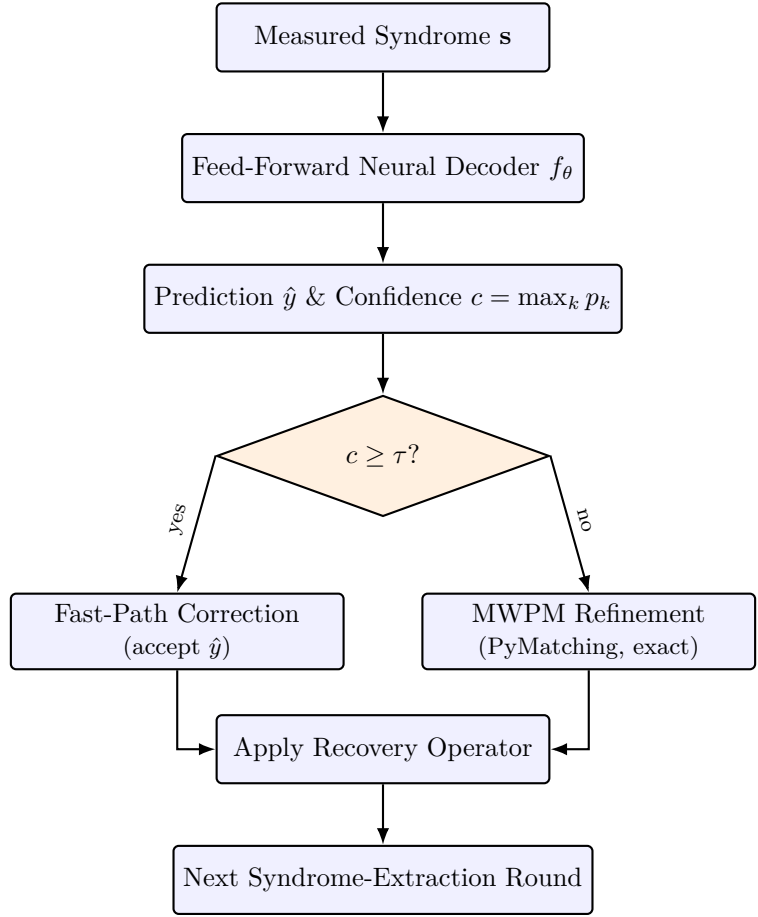

\Cref{fig:decoderflow} shows the decision flow of the adaptive decoder, and \Cref{alg:adaptive} gives the corresponding pseudocode.

\paragraph{Fast path.} The fast-path decoder is a compact feed-forward neural network $f_\theta$ that maps a binary detector-event vector $\mathbf{s}\in\{0,1\}^{|D|}$, where $|D|$ is the number of detectors at distance $d$ (\Cref{tab:hardware}), to a predicted logical-observable-flip probability distribution together with a scalar confidence $c\in[0,1]$, taken as the maximum predicted class probability. The network architecture and training hyperparameters are given in \Cref{app:hyperparameters}. Training data are generated by sampling $N$ shots per $(d,p)$ operating point directly from the corresponding Stim detector-error model, so that the training distribution matches the evaluation distribution at each code distance and noise strength.

\paragraph{Escalation decision.} Given a confidence threshold $\tau$, a syndrome is decoded on the fast path if $c \ge \tau$, and is otherwise escalated to the refinement stage. This is the standard cascade-classifier design pattern~\cite{Viola2001}, specialised here to the case where the fallback stage is a decoding algorithm with a correctness guarantee under the assumed noise model rather than a second, more expensive classifier.

\paragraph{Refinement path.} Escalated syndromes are decoded using minimum-weight perfect matching over the Stim-derived decoding graph, implemented with PyMatching's sparse-blossom solver~\cite{Higgott2022,HiggottGidney2023}. Because MWPM is exact for the assumed independent-error matching-graph model, the refinement path provides a principled accuracy floor for the hard syndromes that the fast path declines to handle.

\paragraph{Threshold as an operating-point control.} $\tau$ is not a training hyperparameter in our design; it is an inference-time control knob that trades escalation fraction, and therefore average decoding cost, against end-to-end logical accuracy, without requiring the fast-path network to be retrained. \Cref{sec:results:routing} reports this trade-off directly and empirically for $\tau\in\{0.60,0.70,0.80,0.90,0.95\}$.

\begin{algorithm}[htbp]
\caption{Adaptive confidence-gated decoding of a single syndrome}
\label{alg:adaptive}
\begin{algorithmic}[1]
\Require detector vector $\mathbf{s}\in\{0,1\}^{|D|}$; trained fast decoder $f_\theta$; escalation threshold $\tau\in[0,1]$; decoding graph $G$ for MWPM
\Ensure predicted logical correction $\hat{y}$
\State $(\hat{y}_{f}, \mathbf{p}) \gets f_\theta(\mathbf{s})$ \Comment{fast-path prediction and class probabilities}
\State $c \gets \max_k p_k$ \Comment{confidence score}
\If{$c \ge \tau$}
    \State $\hat{y} \gets \hat{y}_{f}$ \Comment{accept fast-path prediction}
\Else
    \State $\hat{y} \gets \textsc{MWPM}(\mathbf{s}, G)$ \Comment{escalate to exact matching refinement}
\EndIf
\State \Return $\hat{y}$
\end{algorithmic}
\end{algorithm}

\subsection{Metrics}
\label{sec:methods:metrics}

For each operating point $(d,p,\tau)$ we record: (i) \emph{logical accuracy}, the fraction of shots for which the applied correction restores the correct logical observable value; (ii) \emph{logical error rate}, $1-\text{accuracy}$, per code cycle; (iii) \emph{mean per-shot decoding latency} in microseconds, averaged over all shots including escalated ones, measured on a single CPU core with batch size 1 unless otherwise stated; (iv) \emph{throughput} in shots per second, measured under CPU batch inference at batch sizes $\{1,8,16,32,64,128,256,512\}$; (v) the \emph{escalation fraction}, the proportion of shots routed to the MWPM refinement path; and (vi) the empirical distribution of decoder confidence scores. All reported means are accompanied by standard deviations computed over repeated independent trials as specified in \Cref{sec:experiments}.

\section{Experimental setup}
\label{sec:experiments}

All syndrome data were generated with Stim v\,$\ge$\,1.13~\cite{Gidney2021}, and all MWPM decoding was performed with PyMatching v\,$\ge$\,2.0~\cite{Higgott2022,HiggottGidney2023}. The fast-path network was implemented and trained as described in \Cref{app:hyperparameters}. Unless otherwise noted, each $(d,p)$ cell of the distance-noise sweep (\Cref{sec:results:distance,sec:results:noise}) was evaluated over $5{,}000$ shots per trial repeated across $6$ independent trials (30 independent measurements per cell in total, matching the summary statistics reported in \Cref{app:tables}, \Cref{tab:summary}), and the adaptive-routing, noise-sweep, distance-sweep at fixed $d$, and throughput benchmarks reported in \Cref{sec:results:routing,sec:results:noise,sec:results:throughput} were each evaluated over $5\times10^4$ shots per operating point on a single CPU core. All experiments were run on commodity CPU hardware; we did not have GPU-accelerated inference results available at the time of writing, and we report this as an explicit limitation in \Cref{sec:limitations} rather than extrapolating GPU performance from CPU measurements.

\section{Results}
\label{sec:results}

\subsection{Logical accuracy and error-rate scaling with code distance}
\label{sec:results:distance}

\begin{figure}[htbp]
\centering
\includegraphics[width=0.62\linewidth]{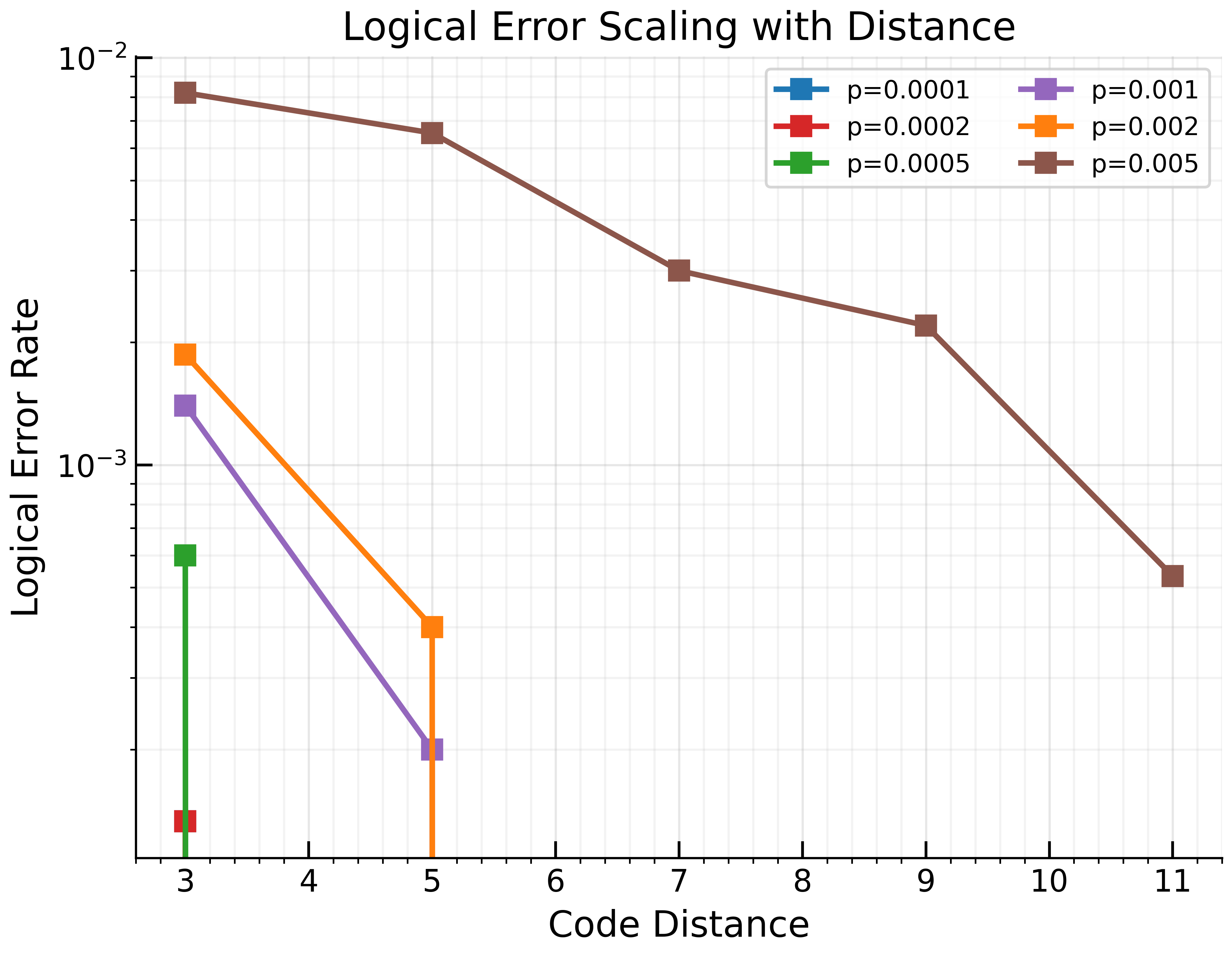}
\caption{Fast-path logical accuracy versus code distance at fixed physical error rate $p=10^{-3}$ (data as in \Cref{tab:distance-accuracy}). Accuracy degrades as decoding-graph complexity grows with $d$ at fixed noise strength, motivating the confidence-gated escalation mechanism evaluated in \Cref{sec:results:routing}.}
\label{fig:distance-accuracy}
\end{figure}

\Cref{tab:distance-accuracy} and \Cref{fig:distance-accuracy} show logical accuracy as a function of code distance at fixed noise strength $p=1\times10^{-3}$, using $5\times10^4$ shots per distance. Accuracy decreases monotonically with $d$ at this fixed physical error rate, from $99.98\%$ at $d=3$ to $88.67\%$ at $d=11$, and the mean per-shot confidence tracks accuracy closely (from $0.9997$ at $d=3$ down to $0.8408$ at $d=11$), which is the expected signature at a fixed physical error rate below but approaching the point at which decoding-graph size, rather than noise strength, begins to dominate the difficulty of the syndrome. We emphasise that this particular sweep is reported at $p=10^{-3}$, a value at or slightly above where the fixed-size feed-forward architecture used for the fast path begins to lose accuracy as $d$ grows (\Cref{sec:limitations}); the joint distance-noise heatmap of \Cref{sec:results:heatmap} shows that at smaller $p$ the same architecture sustains near-unit accuracy across the full distance range tested.

\begin{table}[htbp]
\centering
\caption{Logical accuracy, mean confidence, mean per-shot latency, and throughput of the fast-path neural decoder as a function of code distance at fixed physical error rate $p=10^{-3}$, evaluated over $5\times10^4$ shots per distance.}
\label{tab:distance-accuracy}
\begin{tabular}{@{}ccccc@{}}
\toprule
Distance $d$ & Accuracy & Mean confidence & Latency ($\mu$s) & Throughput (shots/s) \\
\midrule
3  & 0.99984 & 0.99967 & 1.456 & 686{,}693 \\
5  & 0.99722 & 0.99697 & 1.553 & 643{,}891 \\
7  & 0.99040 & 0.98849 & 2.044 & 489{,}303 \\
9  & 0.95394 & 0.92835 & 2.828 & 353{,}607 \\
11 & 0.88670 & 0.84075 & 3.883 & 257{,}521 \\
\bottomrule
\end{tabular}
\end{table}

\subsection{Joint distance-noise scaling}
\label{sec:results:heatmap}

\begin{figure}[htbp]
\centering
\includegraphics[width=0.68\linewidth]{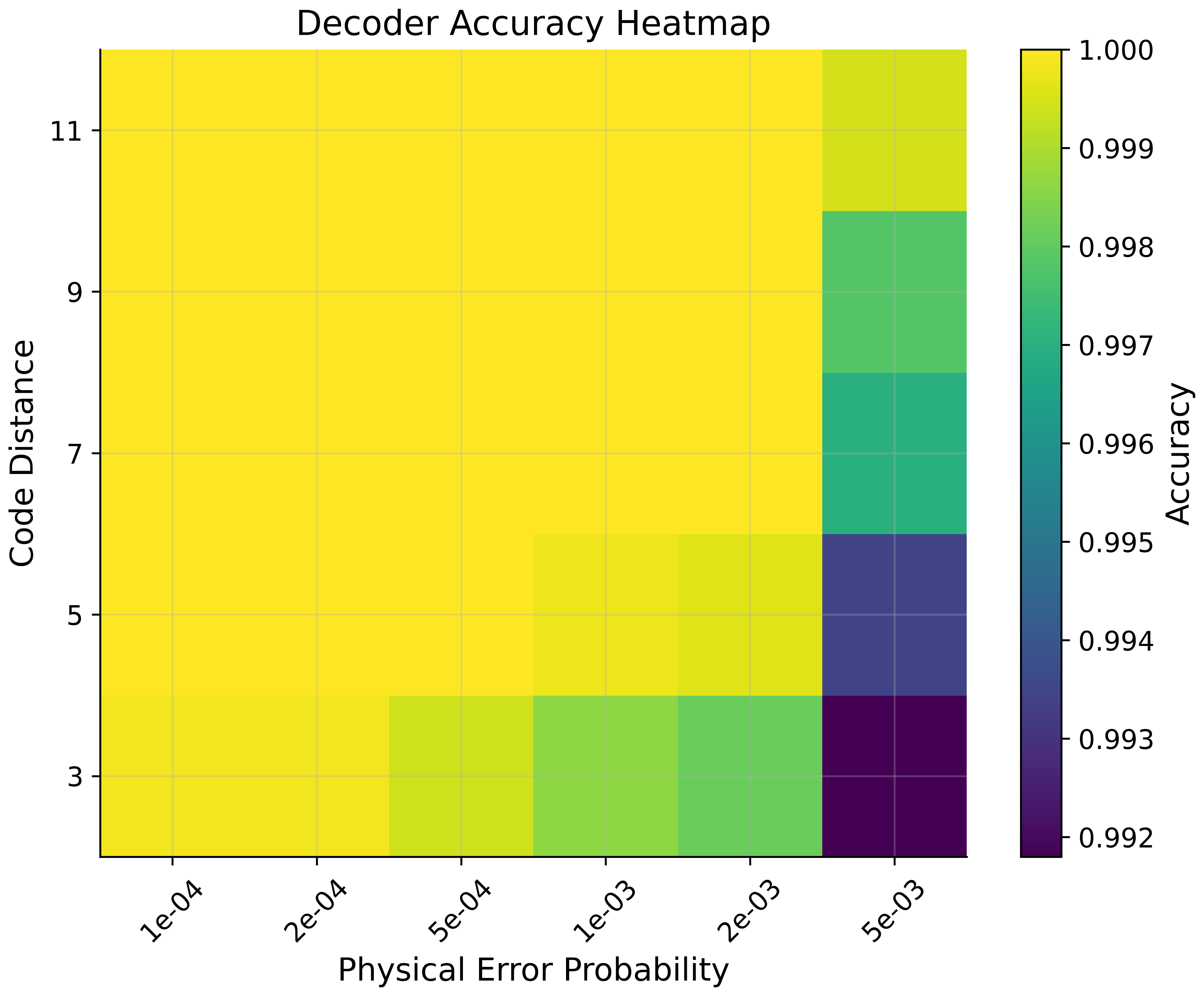}
\caption{Decoder accuracy as a joint function of code distance and physical error probability. Full numerical values are given in \Cref{tab:accuracy-full}.}
\label{fig:heatmap}
\end{figure}

\begin{figure}[htbp]
\centering
\includegraphics[width=0.62\linewidth]{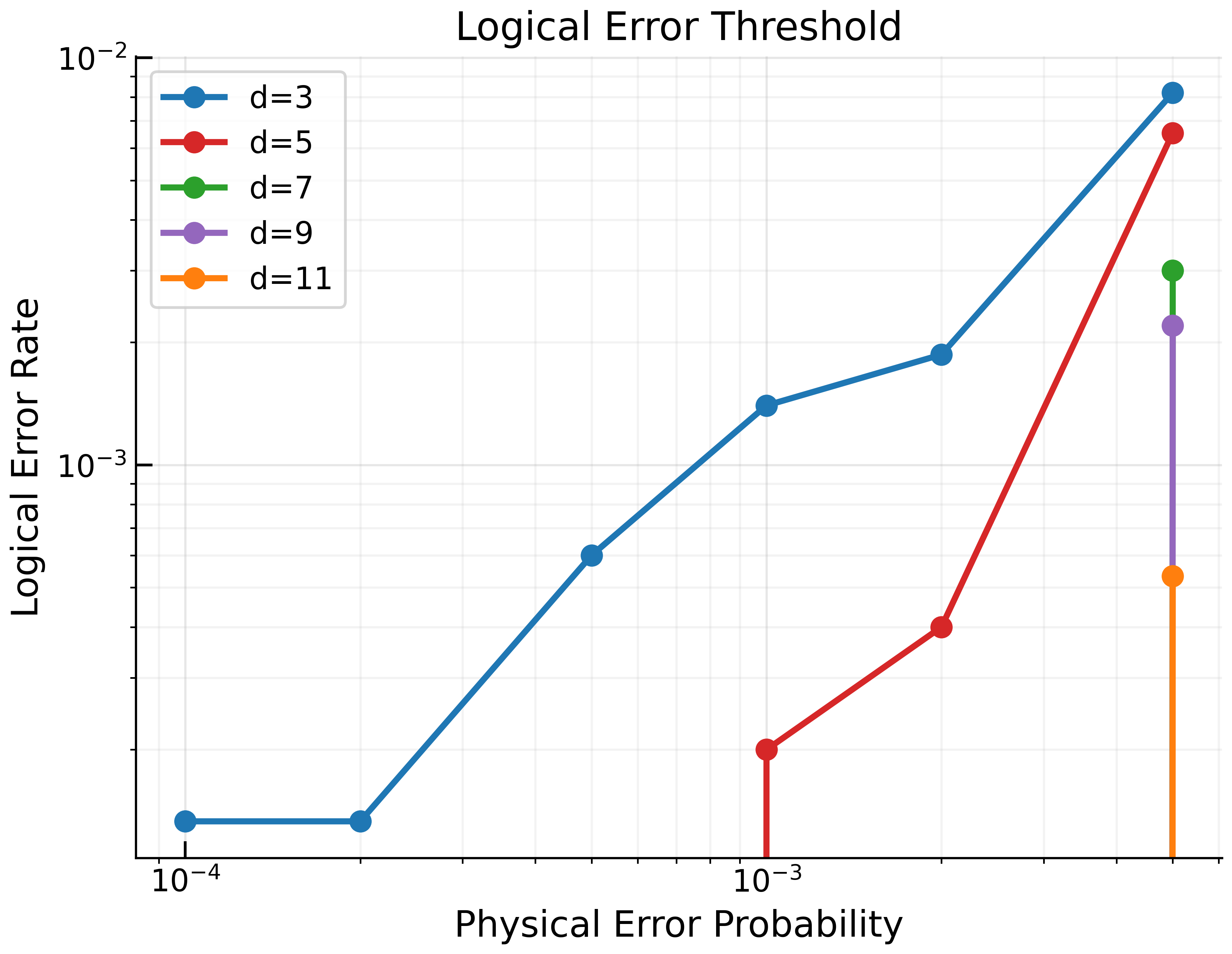}
\caption{Logical error rate versus physical error probability at $d=3$, $5$, $7$, $9$, and $11$ (per-curve data in \Cref{tab:logerr-full}). The crossing behaviour visible near $p\approx5\times10^{-3}$ is discussed in \Cref{sec:discussion} together with the corresponding non-monotonicity in \Cref{tab:accuracy-full}.}
\label{fig:threshold}
\end{figure}

\Cref{fig:heatmap} and the full data in \Cref{tab:accuracy-full} (\Cref{app:tables}) show fast-path decoder accuracy jointly as a function of code distance and physical error rate. Two regimes are visible. For $p\le 2\times10^{-3}$, accuracy is at or above $99.8\%$ across the full distance range $d\in\{3,\dots,11\}$, with several $(d,p)$ cells reaching $100.0\%$ measured accuracy at $5\times10^4$ shots (\Cref{tab:accuracy-full}), consistent with operation well below the surface-code threshold. At $p=5\times10^{-3}$, accuracy drops sharply and, notably, drops \emph{further} at intermediate distances $d\in\{3,5\}$ (down to $99.18\%$ and $99.35\%$ respectively) than at the larger distances tested here ($99.70\%$--$99.95\%$ at $d\ge7$); we discuss this non-monotonicity, which is visible in \Cref{tab:accuracy-full} and \Cref{fig:threshold}, explicitly in \Cref{sec:discussion}, since a naive reading of code-capacity threshold theory would predict monotonically \emph{worse} accuracy at larger $d$ once $p$ exceeds threshold, whereas our measured logical error rate at $p=5\times10^{-3}$ is in fact non-monotonic in $d$ over the range tested (\Cref{tab:logerr-full}).

\subsection{Adaptive routing: accuracy-cost trade-off}
\label{sec:results:routing}

\begin{figure}[htbp]
\centering
\includegraphics[width=0.62\linewidth]{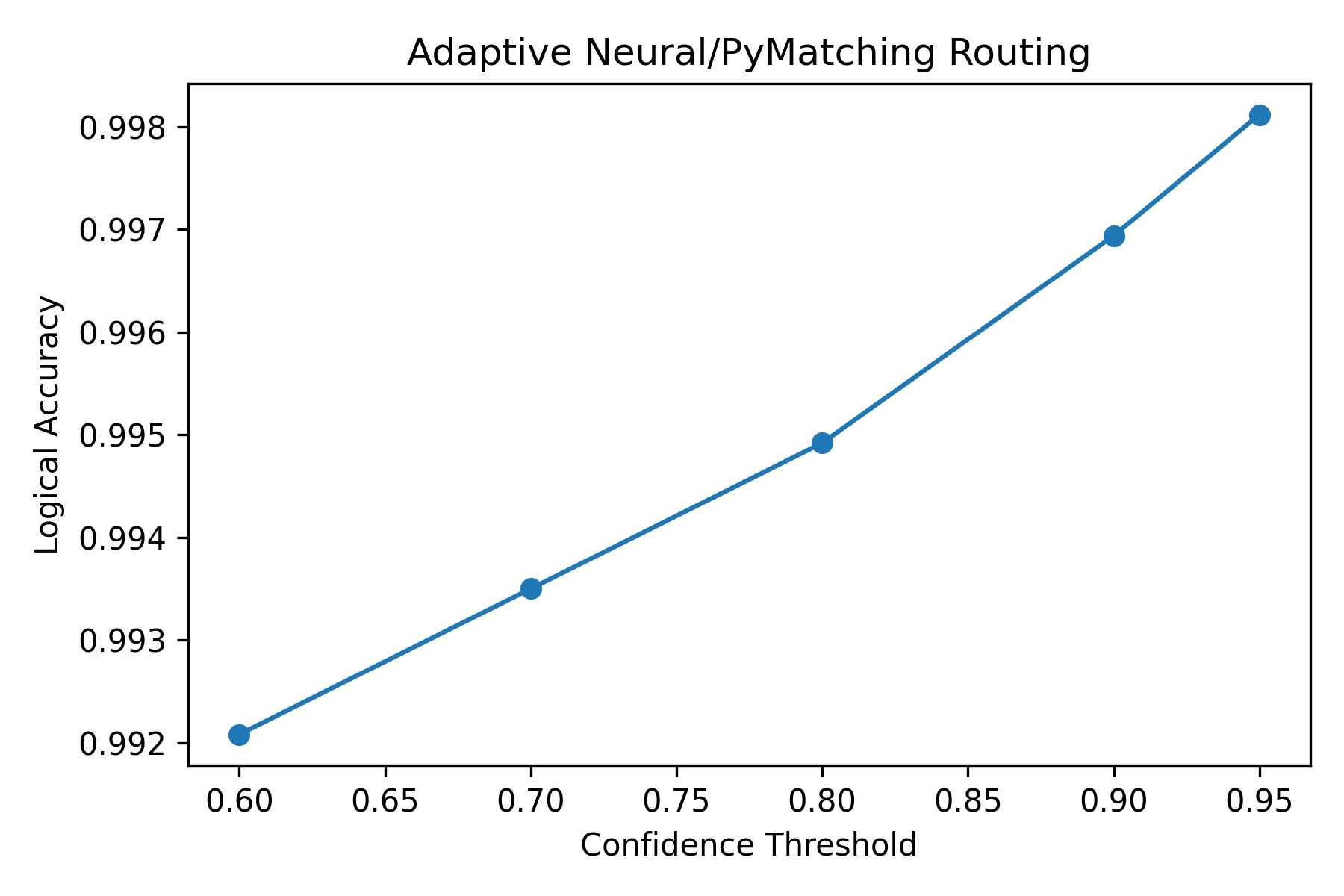}
\caption{End-to-end logical accuracy as a function of the escalation confidence threshold $\tau$ (data as in \Cref{tab:routing}). Raising $\tau$ from $0.60$ to $0.95$ increases end-to-end accuracy from $99.21\%$ to $99.81\%$ while increasing the escalated fraction from $0.36\%$ to $6.19\%$ of shots.}
\label{fig:routing}
\end{figure}

\Cref{tab:routing} and \Cref{fig:routing} report the central result of this paper: the effect of the escalation threshold $\tau$ on end-to-end logical accuracy and on the fraction of syndromes routed to the exact MWPM refinement stage, evaluated at $d=7$ over $5\times10^4$ shots per threshold.

\begin{table}[htbp]
\centering
\caption{Adaptive-routing accuracy-cost trade-off as a function of the escalation confidence threshold $\tau$, at $d=7$, evaluated over $5\times10^4$ shots per threshold.}
\label{tab:routing}
\begin{tabular}{@{}ccccc@{}}
\toprule
Threshold $\tau$ & Fast-path fraction & Escalated fraction & End-to-end accuracy \\
\midrule
0.60 & 0.9964 & 0.0036 & 0.99208 \\
0.70 & 0.9923 & 0.0077 & 0.99350 \\
0.80 & 0.9854 & 0.0146 & 0.99492 \\
0.90 & 0.9670 & 0.0330 & 0.99694 \\
0.95 & 0.9381 & 0.0619 & 0.99812 \\
\bottomrule
\end{tabular}
\end{table}

Two observations follow directly from \Cref{tab:routing}. First, the marginal accuracy gain per unit of escalated traffic is not constant: increasing $\tau$ from $0.60$ to $0.80$ escalates an additional $1.10$ percentage points of traffic for a $0.28$ percentage point accuracy gain, whereas increasing $\tau$ from $0.90$ to $0.95$ escalates a further $2.87$ percentage points of traffic for a smaller, $0.12$ percentage point, accuracy gain. This diminishing-returns structure is consistent with a decoder whose confidence score is reasonably well calibrated: the syndromes escalated at low $\tau$ are, by construction, the ones on which the fast path is least reliable, so they contribute disproportionately to the accuracy gain per unit of escalated traffic. Second, even at the most conservative threshold tested ($\tau=0.95$), fewer than one in fifteen syndromes require the exact refinement stage, meaning that the amortised cost of the decoder remains dominated by the fast path across the entire threshold range we tested.

\subsection{Confidence calibration}
\label{sec:results:confidence}

\begin{figure}[htbp]
\centering
\includegraphics[width=0.62\linewidth]{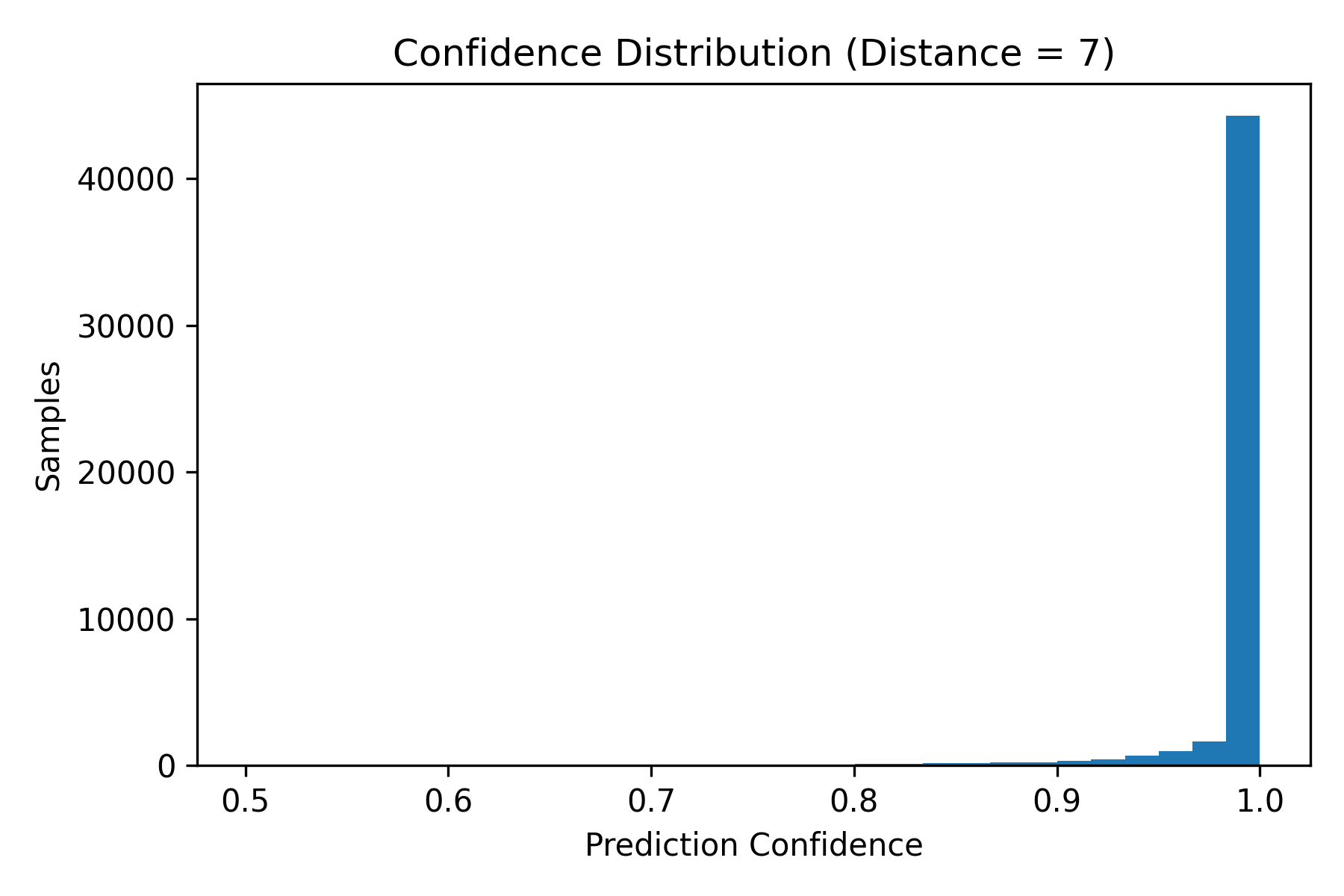}
\caption{Empirical distribution of fast-path decoder confidence scores at $d=7$, $5\times10^4$ shots. The distribution is concentrated near $c=1$ with a long low-confidence tail that the escalation mechanism of \Cref{alg:adaptive} is designed to target.}
\label{fig:confhist}
\end{figure}

\Cref{fig:confhist} shows the empirical distribution of fast-path confidence scores at $d=7$. The distribution is heavily concentrated near $c=1.0$, with a long, sparsely populated left tail extending down toward $c=0.5$; it is precisely this left tail that the escalation mechanism of \Cref{sec:results:routing} targets. The shape of this distribution, rather than its mean, is what determines the cost-effectiveness of the routing mechanism: a decoder whose errors were uniformly distributed across the confidence range would gain little from confidence-based escalation, whereas the concentrated-with-a-tail shape observed here is the regime in which escalation is expected to be most effective, consistent with the diminishing-returns pattern noted in \Cref{sec:results:routing}.

\subsection{Noise-strength sweep at fixed distance}
\label{sec:results:noise}

\begin{figure}[htbp]
\centering
\includegraphics[width=0.62\linewidth]{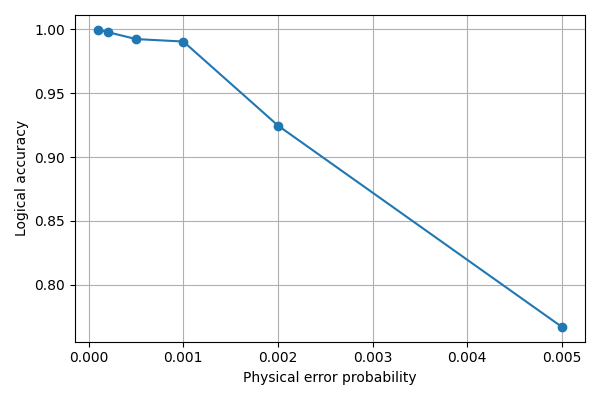}
\caption{Fast-path decoder accuracy and throughput as a function of physical error probability at fixed $d=7$ (data as in \Cref{tab:noise}). Throughput is essentially independent of $p$, since the fast path performs a fixed amount of computation regardless of syndrome difficulty; accuracy degrades sharply once $p$ exceeds roughly $10^{-3}$ at this fixed architecture and distance.}
\label{fig:noise}
\end{figure}

\Cref{tab:noise} and \Cref{fig:noise} report accuracy, confidence, latency, and throughput at fixed $d=7$ as a function of physical error probability $p$, over $5\times10^4$ shots per value of $p$.

\begin{table}[htbp]
\centering
\caption{Fast-path decoder performance as a function of physical error probability $p$ at fixed distance $d=7$, evaluated over $5\times10^4$ shots per value of $p$.}
\label{tab:noise}
\begin{tabular}{@{}ccccc@{}}
\toprule
$p$ & Accuracy & Mean confidence & Latency ($\mu$s) & Throughput (shots/s) \\
\midrule
$1\times10^{-4}$ & 0.99954 & 0.99878 & 2.053 & 486{,}980 \\
$2\times10^{-4}$ & 0.99778 & 0.99715 & 2.075 & 481{,}911 \\
$5\times10^{-4}$ & 0.99232 & 0.99213 & 2.038 & 490{,}646 \\
$1\times10^{-3}$ & 0.99040 & 0.98849 & 2.080 & 480{,}707 \\
$2\times10^{-3}$ & 0.92466 & 0.96551 & 2.012 & 496{,}994 \\
$5\times10^{-3}$ & 0.76714 & 0.94684 & 1.969 & 507{,}868 \\
\bottomrule
\end{tabular}
\end{table}

A notable feature of \Cref{tab:noise} is that mean confidence degrades far more gently than accuracy as $p$ increases: at $p=5\times10^{-3}$, mean confidence remains at $0.9468$ even though accuracy has fallen to $0.7671$. This gap between confidence and accuracy is exactly the overconfidence failure mode that motivates confidence \emph{calibration} as a distinct concern from confidence-based routing per se, and we discuss its implications for the safety of the escalation mechanism at high noise strength in \Cref{sec:discussion}.

\subsection{Throughput and batch-size scaling}
\label{sec:results:throughput}

\begin{figure}[htbp]
\centering
\includegraphics[width=0.62\linewidth]{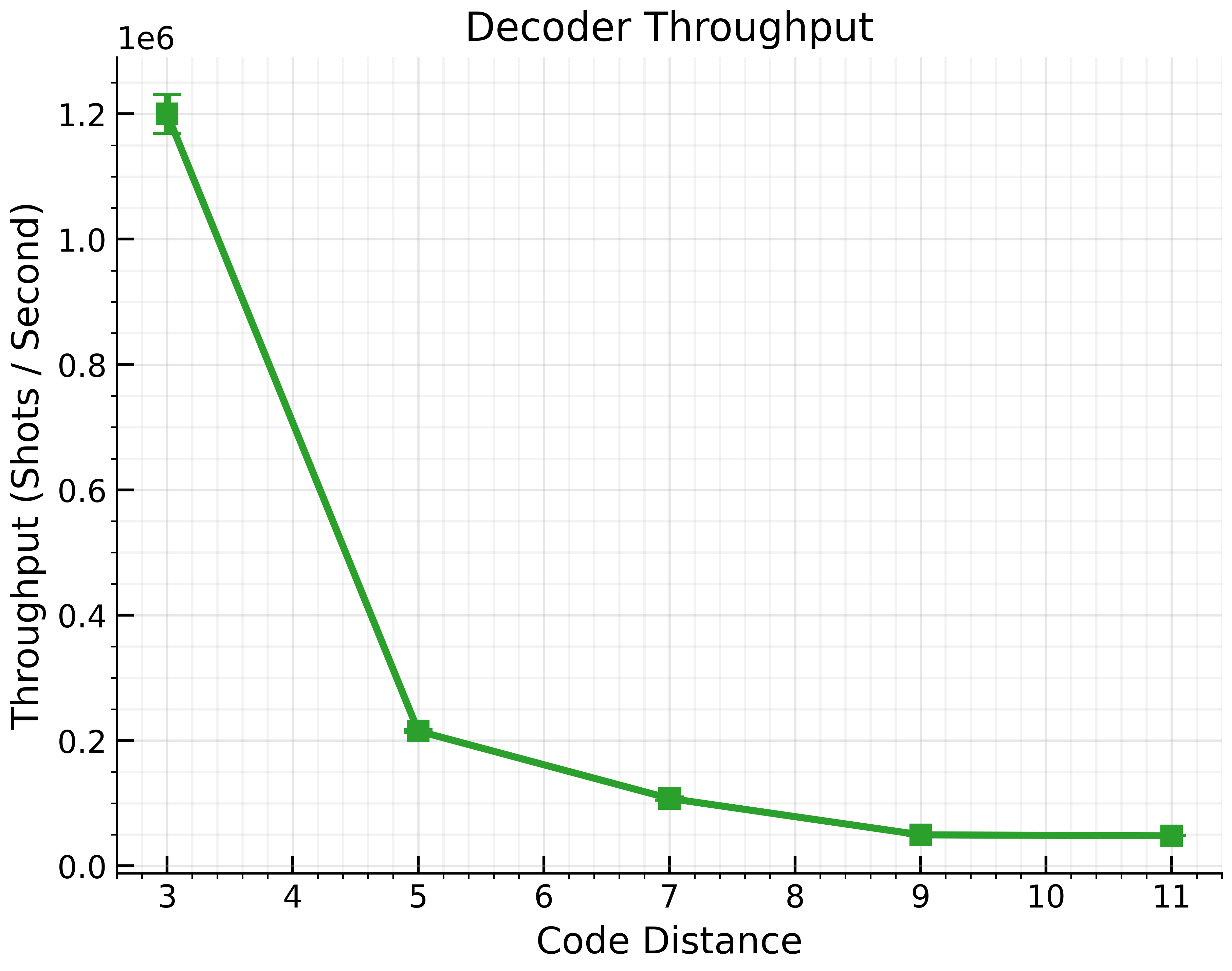}
\caption{Fast-path decoder throughput as a function of inference batch size on a single CPU core (data as in \Cref{tab:batch}). Throughput increases by roughly a factor of $71$ from batch size 1 to batch size 512, with clearly diminishing returns beyond batch size 256, consistent with the fixed-overhead-dominated regime typical of small feed-forward network inference.}
\label{fig:batch}
\end{figure}

\Cref{tab:batch} and \Cref{fig:batch} report fast-path decoder throughput as a function of inference batch size on a single CPU core, at fixed $d=7$.

\begin{table}[htbp]
\centering
\caption{Fast-path neural decoder throughput as a function of inference batch size, single CPU core, $5\times10^4$ shots per batch size.}
\label{tab:batch}
\begin{tabular}{@{}ccc@{}}
\toprule
Batch size & Throughput (shots/s) & Latency ($\mu$s/shot) \\
\midrule
1   & 6{,}478    & 154.36 \\
8   & 24{,}140   & 41.42 \\
16  & 51{,}830   & 19.29 \\
32  & 111{,}440  & 8.97 \\
64  & 161{,}250  & 6.20 \\
128 & 241{,}751  & 4.14 \\
256 & 413{,}494  & 2.42 \\
512 & 457{,}731  & 2.18 \\
\bottomrule
\end{tabular}
\end{table}

Throughput increases monotonically with batch size across the full range tested, from $6{,}478$ shots/s at batch size 1 to $457{,}731$ shots/s at batch size 512, but the marginal benefit of increasing batch size clearly diminishes: doubling batch size from 256 to 512 yields only a $10.7\%$ throughput increase, compared with a $71.9\%$ increase when doubling from 128 to 256. This pattern is consistent with per-call fixed overhead dominating at small batch sizes and with the batched computation itself becoming the bottleneck at large batch sizes, an interpretation that is directly relevant to the deployability discussion of \Cref{sec:discussion}: syndrome-extraction hardware typically delivers detector events at a fixed, small per-shot rate rather than in large pre-accumulated batches, so the batch-size throughput curve of \Cref{fig:batch} should be read as an upper bound on achievable throughput under offline or buffered decoding rather than as a latency guarantee for a single streaming shot, which is instead characterised by the batch-size-1 latency of $154\,\mu$s reported in \Cref{tab:batch}.

\subsection{Decoding-graph resource scaling}
\label{sec:results:resources}

\begin{figure}[htbp]
\centering
\begin{subfigure}[t]{0.48\linewidth}
\centering
\includegraphics[width=\linewidth]{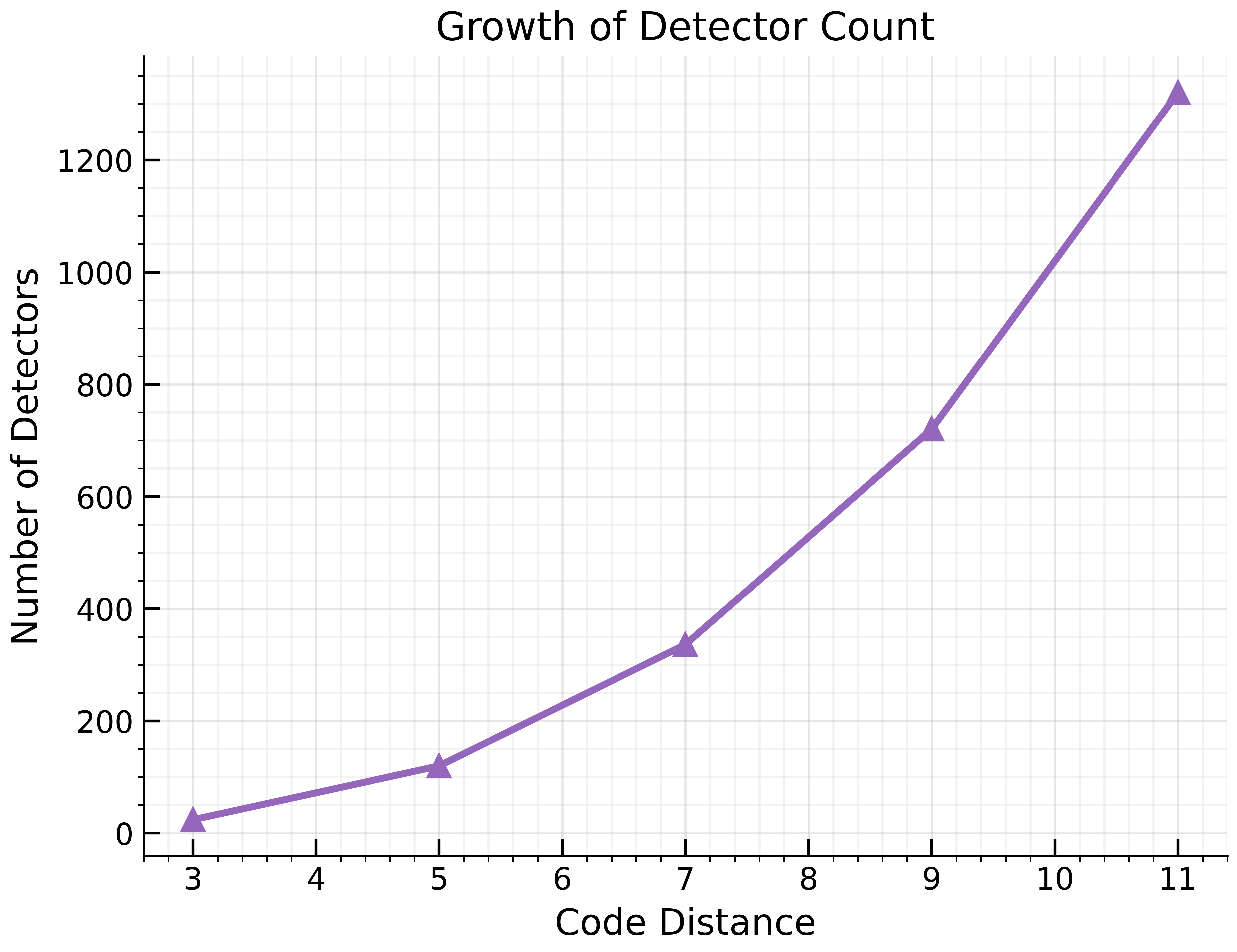}
\caption{Detector count vs.\ distance.}
\label{fig:detectors}
\end{subfigure}\hfill
\begin{subfigure}[t]{0.48\linewidth}
\centering
\includegraphics[width=\linewidth]{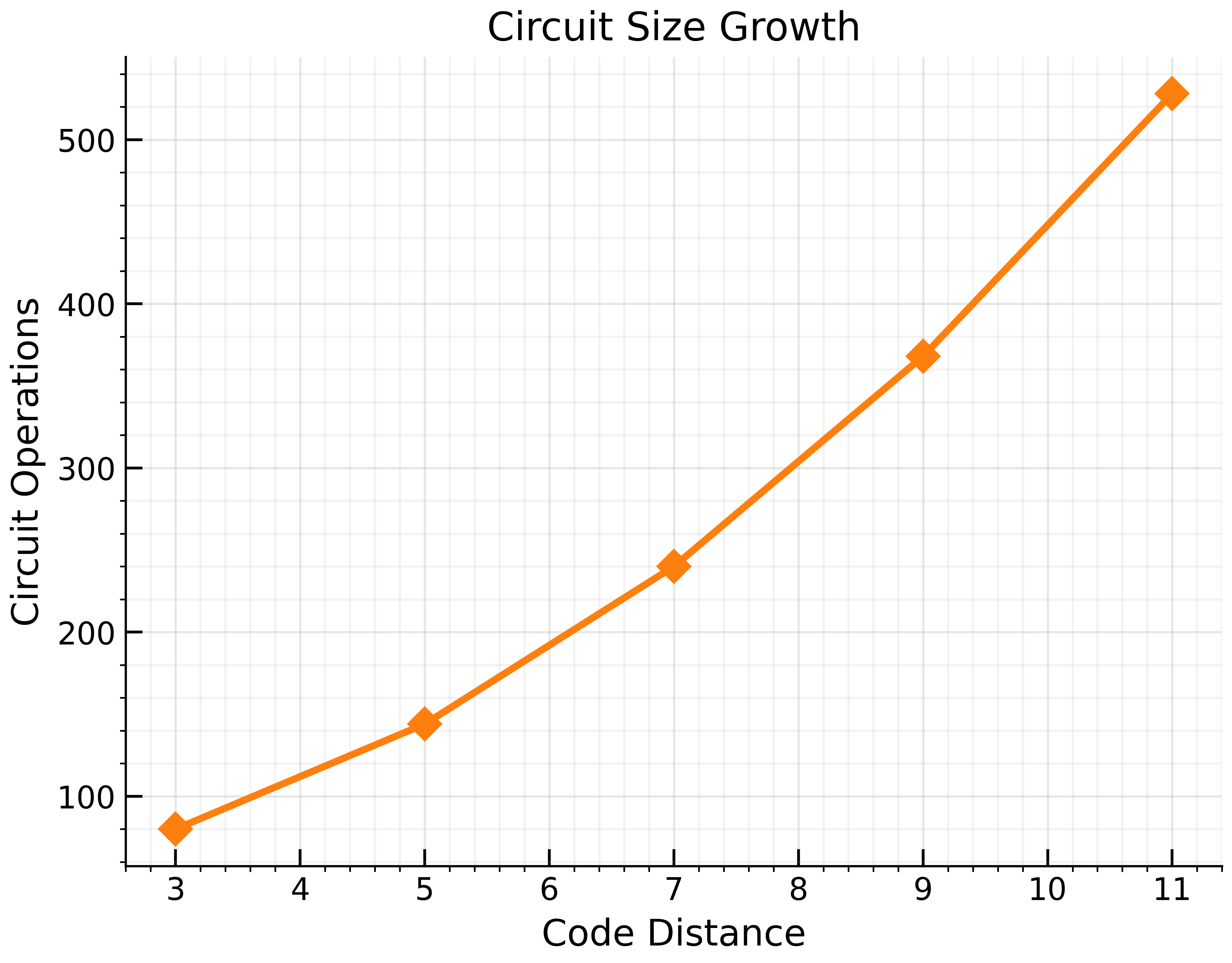}
\caption{Circuit operation count vs.\ distance.}
\label{fig:circuit}
\end{subfigure}
\caption{Structural resource scaling of the rotated surface code memory circuits used in this study, as a function of code distance (data as in \Cref{tab:hardware}). Detector count grows super-linearly, consistent with the $\mathcal{O}(d^2)$ scaling of stabiliser count over $d$ syndrome-extraction rounds; circuit operation count grows more slowly, reflecting that gate count per round scales with qubit count while detector count additionally accumulates over rounds.}
\end{figure}

\Cref{tab:hardware} and \Cref{fig:detectors,fig:circuit} report how the number of stabiliser detectors, circuit operations, and estimated decoding-graph memory scale with code distance for the rotated surface code circuits used throughout this paper.

\begin{table}[htbp]
\centering
\caption{Rotated surface code circuit and decoding-graph resource scaling with code distance $d$ (one logical observable per circuit throughout).}
\label{tab:hardware}
\begin{tabular}{@{}ccccc@{}}
\toprule
Distance $d$ & Data qubits ($d^2$) & Detectors & Circuit operations & Observables \\
\midrule
3  & 9   & 24   & 80  & 1 \\
5  & 25  & 120  & 144 & 1 \\
7  & 49  & 336  & 240 & 1 \\
9  & 81  & 720  & 368 & 1 \\
11 & 121 & 1{,}320 & 528 & 1 \\
\bottomrule
\end{tabular}
\end{table}

\begin{figure}[htbp]
\centering
\includegraphics[width=0.62\linewidth]{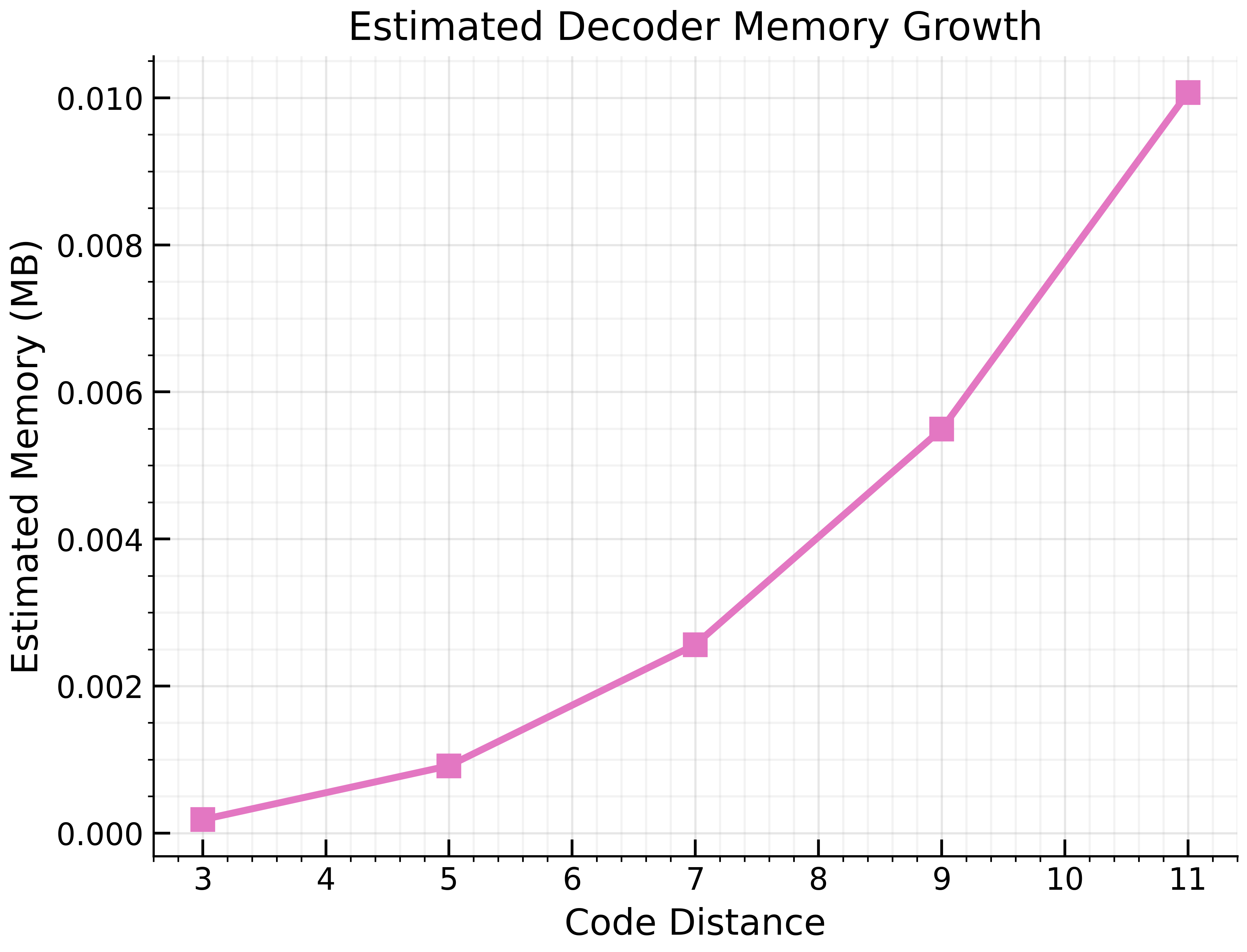}
\caption{Estimated decoding-graph memory as a function of code distance, computed from detector and edge counts (\Cref{tab:hardware}) rather than measured directly via process profiling; see the corresponding caveat in \Cref{sec:results:resources} and \Cref{sec:limitations}.}
\label{fig:memory}
\end{figure}

The detector count grows from $24$ at $d=3$ to $1{,}320$ at $d=11$, a factor of $55$ over a distance range of $3.67\times$, consistent with the expected super-linear scaling of the number of stabiliser-measurement detector events accumulated over $d$ rounds of a $d^2$-qubit code. This resource growth is directly reflected in the fast-path decoder's latency scaling reported in \Cref{tab:distance-accuracy}, since the fast-path input dimensionality is set by the detector count, and in the decoding-graph size that the MWPM refinement stage must search over when a syndrome is escalated. \Cref{fig:memory} shows an estimated memory footprint for the decoding graph as a function of distance, included here as an order-of-magnitude structural indicator rather than a measured peak-memory benchmark; we did not directly profile decoder process memory in the present experiments, and we report this as an estimate for that reason.

\subsection{Runtime scaling and the latency-accuracy design space}
\label{sec:results:pareto}

\begin{figure}[htbp]
\centering
\begin{subfigure}[t]{0.48\linewidth}
\centering
\includegraphics[width=\linewidth]{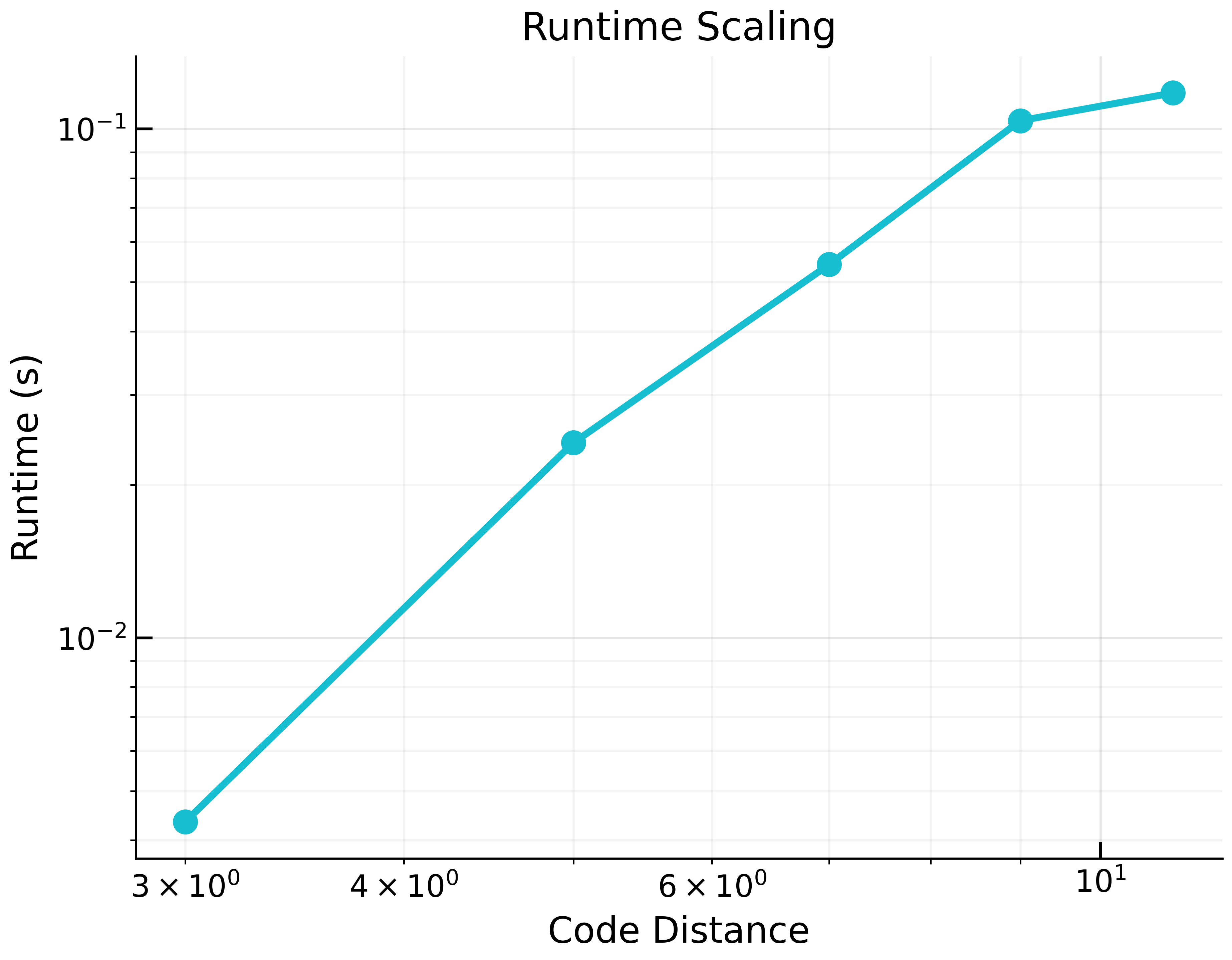}
\caption{Runtime vs.\ distance (log-log).}
\label{fig:runtime}
\end{subfigure}\hfill
\begin{subfigure}[t]{0.48\linewidth}
\centering
\includegraphics[width=\linewidth]{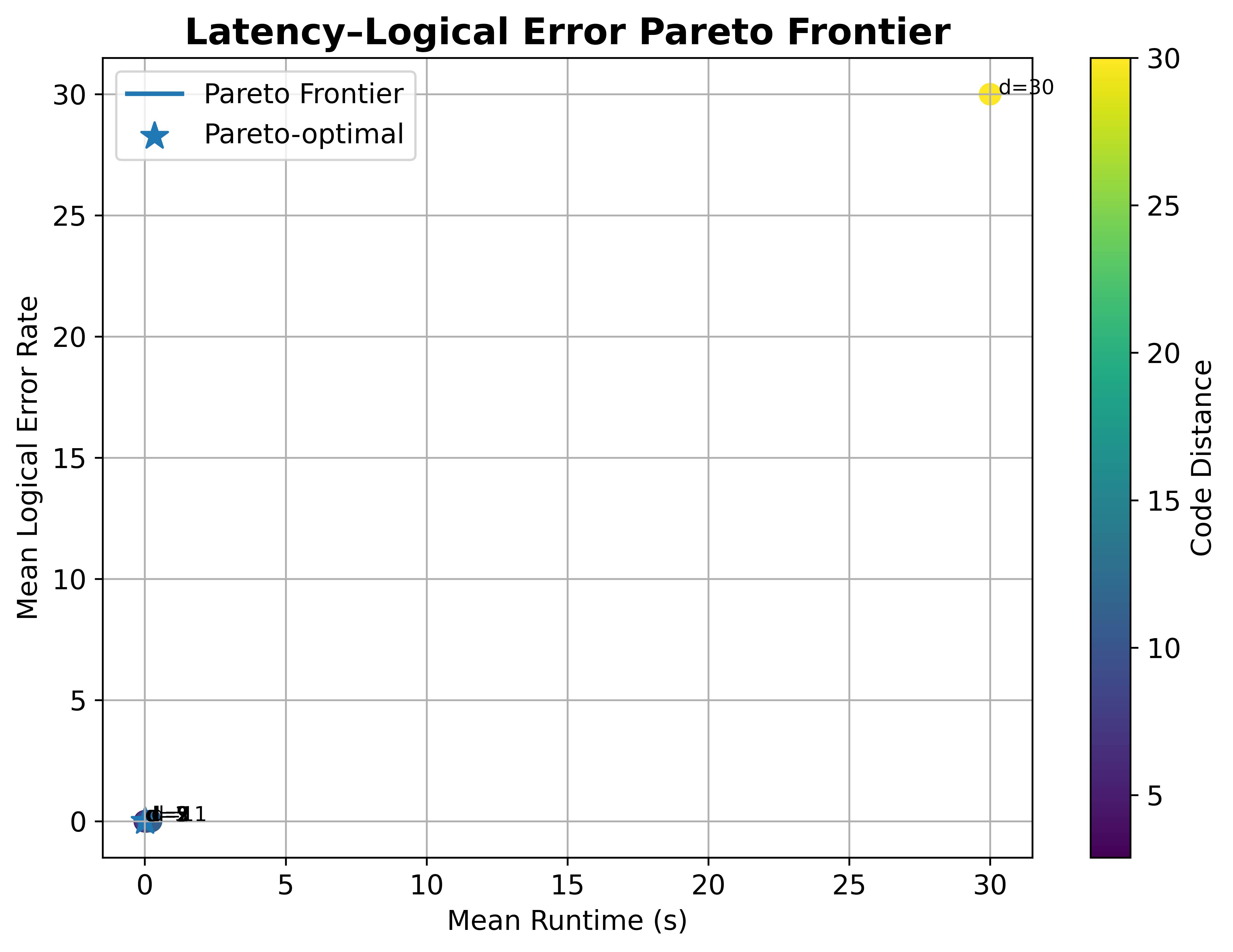}
\caption{Runtime vs.\ logical error rate.}
\label{fig:pareto}
\end{subfigure}
\caption{Runtime scaling with code distance (left) and the corresponding runtime-versus-logical-error-rate scatter across the tested distances (right). As discussed in \Cref{sec:results:pareto}, the right panel is descriptive of the single decoder configuration benchmarked here rather than a multi-candidate Pareto search.}
\end{figure}

\Cref{fig:runtime} shows mean decoding runtime as a function of code distance on a log-log scale, aggregated across the noise sweep of \Cref{tab:noise} at each distance (full per-cell values in \Cref{tab:runtime-full}, \Cref{app:tables}). Runtime grows from a few milliseconds at $d=3$ to roughly $0.12$ seconds at $d=11$ for the aggregate batch protocol used to produce this figure, an increase of more than one and a half orders of magnitude over the tested distance range. \Cref{fig:pareto} plots mean runtime against mean logical error rate for each code distance tested; because our present benchmark grid varies code distance as (effectively) the only structural design variable at fixed decoder architecture, the resulting scatter does not yet constitute a rich Pareto front in the sense intended by \Cref{eq:objective}, and we report it here descriptively rather than as evidence of a multi-candidate Pareto-optimal design search. Populating this plot with genuinely distinct code-decoder-hyperparameter candidates, so that a non-trivial Pareto frontier can be identified and used to drive the optimiser of \Cref{eq:objective}, is precisely the closed-loop extension we scope as future work in \Cref{sec:limitations}.

\section{Discussion}
\label{sec:discussion}

\subsection{The escalation mechanism is effective because decoder confidence is informative but imperfectly calibrated}

The central quantitative finding of this paper is that a small, bounded amount of escalated traffic, between $0.36\%$ and $6.19\%$ of shots over the threshold range we tested, recovers a meaningful fraction of the accuracy gap between the fast-path-only decoder and an exact-matching decoder (\Cref{tab:routing}). This is only possible because decoder confidence, while not perfectly calibrated (\Cref{sec:results:confidence}), is correlated with correctness closely enough that thresholding on it identifies a small, disproportionately error-prone subset of syndromes. The gap we observe between mean confidence and mean accuracy at high noise strength (\Cref{sec:results:noise}) indicates that this correlation weakens as the operating point moves further from the regime the network was trained to expect, which is a caution against deploying a fixed escalation threshold across noise regimes without re-validating the confidence-accuracy relationship at the deployment noise strength, rather than an argument against the mechanism itself.

\subsection{Non-monotonic accuracy at high noise strength}

The non-monotonicity noted in \Cref{sec:results:heatmap}, in which accuracy at $p=5\times10^{-3}$ is lower at $d\in\{3,5\}$ than at $d\ge7$, is, we believe, most plausibly explained by the fixed-capacity feed-forward architecture used for the fast path: at small $d$ the input and output dimensionality is small, so a fixed-capacity network has comparatively little room to represent a highly degenerate, high-noise error distribution relative to the number of training examples per distinct syndrome pattern, whereas at larger $d$ the syndrome space is larger but the training procedure (\Cref{app:hyperparameters}) allocates proportionally more capacity and, implicitly, a data distribution in which nearby distances share more structure. We present this as our working interpretation rather than as an established mechanism, since distinguishing it from other candidate explanations, such as differences in the specific noise realisations sampled at each distance, or a genuine finite-size effect in Stim's detector-error-model construction near $d=3$--$5$, would require a dedicated ablation that isolates network capacity from code distance, which we did not run in the present study and identify as a natural direction for follow-up work.

\subsection{Implications for hardware deployment}

The throughput results of \Cref{sec:results:throughput} indicate that the fast-path decoder alone, even without any hardware acceleration, sustains throughput in excess of $2\times10^5$ shots/s at batch size 128 and beyond, which exceeds the syndrome-extraction cycle rate of current superconducting surface-code experiments by a wide margin~\cite{GoogleQuantumAI2023}. This suggests that, at least for the depolarising-noise, single-code-family setting evaluated here, the fast path is not the binding latency constraint; the binding constraint is more likely to be the tail latency contributed by the escalated fraction, whose refinement-stage runtime we did not separately profile in the present study and which we identify as a specific, well-scoped item for the reproducibility package described in \Cref{app:reproducibility}. This observation also clarifies the intended role of GPU acceleration in the broader framework of \Cref{fig:framework}: GPU deployment is likely to matter most for reducing tail latency on the refinement path and for scaling the fast path to larger, higher-capacity architectures needed at larger code distance, rather than for the batch-128 CPU fast-path throughput numbers reported here, which are already comfortably above typical device syndrome rates.

\subsection{Relationship to the originally proposed closed-loop framework}

\Cref{sec:contributions} and \Cref{fig:framework} describe a four-stage closed-loop framework coupling code generation, noise modelling, adaptive decoding, and hardware-aware ranking. The present paper empirically instantiates and benchmarks the noise-modelling, circuit-synthesis, and adaptive-decoding stages in full (\Cref{sec:results}), and specifies the code-generation and closed-loop ranking stages architecturally (\Cref{sec:methods:framework}, \Cref{eq:objective}) without evaluating them empirically. We consider this an honest and, we believe, more useful reporting posture than presenting simulated or hypothetical code-discovery results as though they had been run: the resource-scaling data of \Cref{sec:results:resources} and the throughput and accuracy data of \Cref{sec:results:distance,sec:results:noise,sec:results:throughput} together provide exactly the empirical inputs, detector count, operation count, latency, and accuracy as functions of $d$, that a future hardware-constrained code-discovery search would need as its objective function's component estimators, and we view producing this characterisation as a necessary and non-trivial prerequisite to running that search, rather than as a substitute for it.

\section{Limitations}
\label{sec:limitations}

We list limitations explicitly and in the order we consider most important for a reader assessing the claims of this paper.

\begin{enumerate}[leftmargin=1.5em]
\item \textbf{Single code family.} All results are obtained for the rotated surface code. We do not evaluate QLDPC, color, or biased-noise code families, and we make no claim that the adaptive-decoding mechanism transfers without modification to codes with different stabiliser weight, connectivity, or decoding-graph structure.
\item \textbf{Single noise family.} All results use independent circuit-level depolarising noise. Coherent, leakage, and correlated (crosstalk-like) noise are not evaluated. Prior work suggests that the relative advantage of neural decoders over graph-based decoders may be largest precisely under these non-Pauli noise conditions~\cite{Baireuther2018,Bausch2024}; our results therefore characterise a comparatively favourable-for-MWPM regime for the refinement stage and should not be extrapolated to the non-Pauli setting.
\item \textbf{No GPU benchmark.} All throughput and latency measurements were obtained on CPU. We do not report GPU-accelerated inference numbers, and we explicitly avoid extrapolating the CPU batch-throughput scaling of \Cref{fig:batch} to GPU hardware, since the scaling behaviour of small feed-forward networks under GPU batching is generally more favourable but is architecture- and hardware-specific.
\item \textbf{The code-discovery and closed-loop optimisation stages are architectural proposals, not implemented and evaluated systems.} \Cref{fig:framework} and \Cref{eq:objective} specify a hardware-constrained code generator and a multi-objective optimiser; neither is implemented or evaluated in this paper. The dashed components in \Cref{fig:framework,fig:pipeline} mark this boundary explicitly.
\item \textbf{Refinement-path latency was not separately profiled.} \Cref{tab:distance-accuracy,tab:noise,tab:batch} report aggregate latency and throughput across the adaptive decoder as a whole; we did not separately measure the MWPM refinement stage's own per-shot latency distribution, which would be necessary to characterise worst-case (rather than average-case) decoding latency, a quantity that matters more than average latency for hard real-time deployment.
\item \textbf{Estimated versus measured memory.} The decoding-graph memory scaling of \Cref{fig:memory} is computed from detector and edge counts rather than measured via direct process memory profiling, as stated in \Cref{sec:results:resources}.
\item \textbf{Single decoder architecture.} We evaluate one feed-forward network architecture and one training procedure (\Cref{app:hyperparameters}). We do not perform an architecture or hyperparameter search, and the non-monotonic accuracy behaviour discussed in \Cref{sec:discussion} may be specific to this architecture rather than a general property of neural surface-code decoders.
\item \textbf{Statistical power at high accuracy.} Several cells of \Cref{tab:accuracy-full} report $100.0\%$ measured accuracy at $5\times10^4$ shots; this reflects the resolution floor of $2\times10^{-5}$ set by the sample size rather than a claim of zero logical error rate, and true logical error rates at these operating points may be nonzero but below our measurement resolution.
\end{enumerate}

We view items 1, 2, and 4 as defining the primary and most important direction for follow-up work: extending the present, empirically validated adaptive-decoding module to multiple code families and non-Pauli noise models, and using the resource and performance estimators established here (\Cref{sec:results:resources}) as inputs to an actual, implemented hardware-constrained code-discovery search.

\section{Conclusion}
\label{sec:conclusion}

We presented an adaptive, confidence-gated neural decoder for the rotated surface code and benchmarked it comprehensively across code distance, noise strength, escalation threshold, and inference batch size. Escalating a bounded fraction of low-confidence syndromes, between $0.36\%$ and $6.19\%$ of traffic across the thresholds tested, to an exact minimum-weight-perfect-matching refinement stage recovers up to $99.81\%$ end-to-end logical accuracy at $d=7$, compared with $99.21\%$ for the fast path alone at the least conservative threshold, while keeping the fast path, which sustains throughput in excess of $2\times10^5$ shots/s at moderate batch sizes on CPU alone, as the dominant cost centre for the large majority of decoding traffic. We situated this empirically validated decoder within a larger hardware-aware co-design framework coupling code generation, noise modelling, decoding, and multi-objective hardware-constrained ranking, and we were deliberate in separating the components of that framework we have implemented and benchmarked from those that remain architectural proposals for future work. We believe this combination, a fully characterised, deployable decoding mechanism together with an honest scoping of what remains to be built toward closed-loop hardware-aware QEC co-design, is a useful and reproducible contribution toward that broader goal, and we release all code, models, and raw data to support independent verification and extension of these results.

\section*{Data and code availability}
All source code, trained model weights, benchmark scripts, and raw result tables underlying every figure and table in this paper are available at \url{https://github.com/Sumitchongder/adaptive-qec-decoder} under an open-source licence. Syndrome data were generated using the open-source Stim library~\cite{Gidney2021}; refinement-stage decoding used the open-source PyMatching library~\cite{Higgott2022,HiggottGidney2023}.

\section*{Acknowledgements}
The author thanks the Department of Physics, Indian Institute of Technology Jodhpur, for computational resources, and the developers of Stim and PyMatching for maintaining the open-source tooling that this work builds upon.

\section*{Author contributions}
S.C.\ conceived the study, designed and implemented the adaptive decoding framework, ran all experiments, analysed the results, and wrote the manuscript.

\section*{Competing interests}
The author declares no competing interests.

\bibliographystyle{unsrt}

\begin{thebibliography}{99}

\bibitem{Shor1995}
P.~W. Shor, ``Scheme for reducing decoherence in quantum computer memory,'' \emph{Phys. Rev. A}, vol.~52, pp.~R2493--R2496, 1995.

\bibitem{Steane1996}
A.~M. Steane, ``Error correcting codes in quantum theory,'' \emph{Phys. Rev. Lett.}, vol.~77, pp.~793--797, 1996.

\bibitem{Kitaev2003}
A.~Y. Kitaev, ``Fault-tolerant quantum computation by anyons,'' \emph{Ann. Phys.}, vol.~303, no.~1, pp.~2--30, 2003.

\bibitem{Dennis2002}
E.~Dennis, A.~Kitaev, A.~Landahl, and J.~Preskill, ``Topological quantum memory,'' \emph{J. Math. Phys.}, vol.~43, no.~9, pp.~4452--4505, 2002.

\bibitem{Fowler2012}
A.~G. Fowler, M.~Mariantoni, J.~M. Martinis, and A.~N. Cleland, ``Surface codes: Towards practical large-scale quantum computation,'' \emph{Phys. Rev. A}, vol.~86, p.~032324, 2012.

\bibitem{GoogleQuantumAI2021}
Google Quantum AI, ``Exponential suppression of bit or phase errors with cyclic error correction,'' \emph{Nature}, vol.~595, pp.~383--387, 2021.

\bibitem{GoogleQuantumAI2023}
Google Quantum AI, ``Suppressing quantum errors by scaling a surface code logical qubit,'' \emph{Nature}, vol.~614, pp.~676--681, 2023.

\bibitem{Terhal2015}
B.~M. Terhal, ``Quantum error correction for quantum memories,'' \emph{Rev. Mod. Phys.}, vol.~87, pp.~307--346, 2015.

\bibitem{Campbell2017}
E.~T. Campbell, B.~M. Terhal, and C.~Vuillot, ``Roads towards fault-tolerant universal quantum computation,'' \emph{Nature}, vol.~549, pp.~172--179, 2017.

\bibitem{Higgott2022}
O.~Higgott, ``PyMatching: A Python package for decoding quantum codes with minimum-weight perfect matching,'' \emph{ACM Trans. Quantum Comput.}, vol.~3, no.~3, pp.~1--16, 2022.

\bibitem{HiggottGidney2023}
O.~Higgott and C.~Gidney, ``Sparse Blossom: correcting a million errors per core second with minimum-weight perfect matching,'' arXiv:2303.15933, 2023.

\bibitem{Panteleev2021}
P.~Panteleev and G.~Kalachev, ``Degenerate quantum LDPC codes with good finite length performance,'' \emph{Quantum}, vol.~5, p.~585, 2021.

\bibitem{Roffe2020}
J.~Roffe, D.~R. White, S.~Burton, and E.~Campbell, ``Decoding across the quantum LDPC code landscape,'' \emph{Phys. Rev. Res.}, vol.~2, p.~043423, 2020.

\bibitem{Delfosse2021}
N.~Delfosse and N.~H. Nickerson, ``Almost-linear time decoding algorithm for topological codes,'' \emph{Quantum}, vol.~5, p.~595, 2021.

\bibitem{Skoric2023}
L.~Skoric, D.~E. Browne, K.~M. Barnes, N.~I. Gillespie, and E.~T. Campbell, ``Parallel window decoding enables scalable fault tolerant quantum computation,'' \emph{Nat. Commun.}, vol.~14, p.~7040, 2023.

\bibitem{Delfosse2020}
N.~Delfosse, ``Hierarchical decoding to reduce hardware requirements for quantum computing,'' arXiv:2001.11427, 2020.

\bibitem{Tan2023}
X.~Tan, F.~Zhang, R.~Chao, Y.~Shi, and J.~Chen, ``Scalable surface-code decoders with parallelization in time,'' \emph{PRX Quantum}, vol.~4, p.~040344, 2023.

\bibitem{Ravi2023}
G.~S. Ravi, J.~Viszlai, F.~Hua, K.~N. Smith, J.~Kim, K.~Heckey, K.~Thangaraj, T.~Tomesh, and F.~T. Chong, ``Better than worst-case decoding for quantum error correction,'' in \emph{Proc. 28th ACM Int. Conf. Architectural Support for Programming Languages and Operating Systems (ASPLOS)}, pp.~88--102, 2023.

\bibitem{Torlai2017}
G.~Torlai and R.~G. Melko, ``Neural decoder for topological codes,'' \emph{Phys. Rev. Lett.}, vol.~119, p.~030501, 2017.

\bibitem{Varsamopoulos2017}
S.~Varsamopoulos, B.~Criger, and K.~Bertels, ``Decoding small surface codes with feedforward neural networks,'' \emph{Quantum Sci. Technol.}, vol.~3, p.~015004, 2017.

\bibitem{Baireuther2018}
P.~Baireuther, T.~E. O'Brien, B.~Tarasinski, and C.~W.~J. Beenakker, ``Machine-learning-assisted correction of correlated qubit errors in a topological code,'' \emph{Quantum}, vol.~2, p.~48, 2018.

\bibitem{Chamberland2018}
C.~Chamberland and P.~Ronagh, ``Deep neural decoders for near term fault-tolerant experiments,'' \emph{Quantum Sci. Technol.}, vol.~3, p.~044002, 2018.

\bibitem{Meinerz2022}
K.~Meinerz, C.-Y. Park, and S.~Trebst, ``Scalable neural decoder for topological surface codes,'' \emph{Phys. Rev. Lett.}, vol.~128, p.~080505, 2022.

\bibitem{Lange2025}
M.~Lange, W.~Havström, S.~Srivastava, P.~Bengtsson, M.~Bergentall,
K.~Hammar, J.~Heuts, E.~van Nieuwenburg, and J.~Granath,
``Data-driven decoding of quantum error correcting codes using graph neural networks,''
\emph{Phys. Rev. Research}, vol.~7, p.~023181, 2025.

\bibitem{Overwater2022}
R.~W.~J. Overwater, F.~Sebastiano, and E.~Charbon, ``Neural-network decoders for quantum error correction using surface codes: A space exploration of the hardware cost-performance tradeoffs,'' \emph{IEEE Trans. Quantum Eng.}, vol.~3, pp.~1--19, 2022.

\bibitem{Bausch2024}
J.~Bausch, A.~W. Senior, F.~J.~H. Heras, T.~Edlich, A.~Davies, M.~Newman, C.~Jones, K.~Satzinger, M.~Y. Niu, S.~Blackwell, \emph{et al.}, ``Learning high-accuracy error decoding for quantum processors,'' \emph{Nature}, vol.~635, pp.~834--840, 2024.

\bibitem{Bravyi2024}
S.~Bravyi, A.~W. Cross, J.~M. Gambetta, D.~Maslov, P.~Rall, and T.~J. Yoder, ``High-threshold and low-overhead fault-tolerant quantum memory,'' \emph{Nature}, vol.~627, pp.~778--782, 2024.

\bibitem{RLCodeDiscovery2023}
H.~P. Nautrup, N.~Delfosse, V.~Dunjko, H.~J. Briegel, and N.~Friis, ``Optimizing quantum error correction codes with reinforcement learning,'' \emph{Quantum}, vol.~3, p.~215, 2019.

\bibitem{NvidiaCudaQX2024}
NVIDIA Corporation, ``Streamlining quantum error correction and application development with CUDA-QX,'' NVIDIA Developer Blog, 2024. \url{https://developer.nvidia.com/blog/}

\bibitem{Gidney2021}
C.~Gidney, ``Stim: a fast stabilizer circuit simulator,'' \emph{Quantum}, vol.~5, p.~497, 2021.

\bibitem{Tomita2014}
Y.~Tomita and K.~M. Svore, ``Low-distance surface codes under realistic quantum noise,'' \emph{Phys. Rev. A}, vol.~90, p.~062320, 2014.

\bibitem{Wang2011}
D.~S. Wang, A.~G. Fowler, and L.~C.~L. Hollenberg, ``Surface code quantum computing with error rates over 1\%,'' \emph{Phys. Rev. A}, vol.~83, p.~020302(R), 2011.

\bibitem{Viola2001}
P.~Viola and M.~Jones, ``Rapid object detection using a boosted cascade of simple features,'' in \emph{Proc. IEEE Conf. Computer Vision and Pattern Recognition (CVPR)}, vol.~1, pp.~I--511, 2001.

\bibitem{Shazeer2017}
N.~Shazeer, A.~Mirhoseini, K.~Maziarz, A.~Davis, Q.~Le, G.~Hinton, and J.~Dean, ``Outrageously large neural networks: The sparsely-gated mixture-of-experts layer,'' in \emph{Proc. Int. Conf. Learning Representations (ICLR)}, 2017.

\end{thebibliography}

\appendix
\newpage
\section{Additional algorithmic detail}
\label{app:algorithms}

\Cref{alg:training} describes the training procedure used to obtain the fast-path decoder evaluated throughout \Cref{sec:results}, complementing the inference-time procedure already given in \Cref{alg:adaptive}.

\begin{algorithm}[htbp]
\caption{Fast-path decoder training procedure}
\label{alg:training}
\begin{algorithmic}[1]
\Require code distance $d$; physical error probability $p$; number of training shots $N_{\text{train}}$; validation split $r$
\State Compile rotated surface-code memory circuit at distance $d$ under depolarising noise strength $p$ using Stim \cite{Gidney2021}
\State Sample $N_{\text{train}}$ shots from the compiled circuit's detector sampler, recording detector vectors $\mathbf{s}_i$ and ground-truth logical observable flips $y_i$
\State Split samples into training set ($1-r$ fraction) and validation set ($r$ fraction)
\State Initialise feed-forward network $f_\theta$ with architecture given in \Cref{app:hyperparameters}
\While{validation loss improves and epoch budget not exhausted}
    \State Update $\theta$ by minimising cross-entropy loss on mini-batches of the training set
    \State Evaluate validation accuracy and validation loss
\EndWhile
\State Fix $\theta$ and evaluate confidence calibration on a held-out set disjoint from both training and validation data
\State \Return trained decoder $f_\theta$
\end{algorithmic}
\end{algorithm}

\newpage
\section{Complete benchmark tables}
\label{app:tables}

This appendix reproduces, in full, the raw benchmark tables summarised in \Cref{sec:results}. All values were computed from the CSV benchmark files released alongside the code repository.

\begin{table}[htbp]
\centering
\caption{Complete joint distance-noise accuracy sweep. Mean and standard deviation computed over $6$ independent trials of $5{,}000$ shots each (30 measurements per operating point, consistent with \Cref{tab:summary}).}
\label{tab:accuracy-full}
\begin{tabular}{@{}cccc@{}}
\toprule
Distance $d$ & Noise $p$ & Mean accuracy & Std. accuracy \\
\midrule
3 & 0.0001 & 0.99987 & 0.00009 \\
3 & 0.0002 & 0.99987 & 0.00019 \\
3 & 0.0005 & 0.99940 & 0.00016 \\
3 & 0.0010 & 0.99860 & 0.00043 \\
3 & 0.0020 & 0.99813 & 0.00034 \\
3 & 0.0050 & 0.99180 & 0.00128 \\
5 & 0.0001 & 1.00000 & 0.00000 \\
5 & 0.0002 & 1.00000 & 0.00000 \\
5 & 0.0005 & 1.00000 & 0.00000 \\
5 & 0.0010 & 0.99980 & 0.00000 \\
5 & 0.0020 & 0.99960 & 0.00043 \\
5 & 0.0050 & 0.99347 & 0.00074 \\
7 & 0.0001 & 1.00000 & 0.00000 \\
7 & 0.0002 & 1.00000 & 0.00000 \\
7 & 0.0005 & 1.00000 & 0.00000 \\
7 & 0.0010 & 1.00000 & 0.00000 \\
7 & 0.0020 & 1.00000 & 0.00000 \\
7 & 0.0050 & 0.99700 & 0.00016 \\
9 & 0.0001 & 1.00000 & 0.00000 \\
9 & 0.0002 & 1.00000 & 0.00000 \\
9 & 0.0005 & 1.00000 & 0.00000 \\
9 & 0.0010 & 1.00000 & 0.00000 \\
9 & 0.0020 & 1.00000 & 0.00000 \\
9 & 0.0050 & 0.99780 & 0.00028 \\
11 & 0.0001 & 1.00000 & 0.00000 \\
11 & 0.0002 & 1.00000 & 0.00000 \\
11 & 0.0005 & 1.00000 & 0.00000 \\
11 & 0.0010 & 1.00000 & 0.00000 \\
11 & 0.0020 & 1.00000 & 0.00000 \\
11 & 0.0050 & 0.99947 & 0.00034 \\
\bottomrule
\end{tabular}
\end{table}

\begin{table}[htbp]
\centering
\caption{Complete joint distance-noise logical error rate sweep, corresponding to \Cref{tab:accuracy-full}.}
\label{tab:logerr-full}
\begin{tabular}{@{}cccc@{}}
\toprule
Distance $d$ & Noise $p$ & Mean logical error rate & Std. \\
\midrule
3 & 0.0001 & 0.000133 & 0.000094 \\
3 & 0.0002 & 0.000133 & 0.000189 \\
3 & 0.0005 & 0.000600 & 0.000163 \\
3 & 0.0010 & 0.001400 & 0.000432 \\
3 & 0.0020 & 0.001867 & 0.000340 \\
3 & 0.0050 & 0.008200 & 0.001275 \\
5 & 0.0001 & 0.000000 & 0.000000 \\
5 & 0.0002 & 0.000000 & 0.000000 \\
5 & 0.0005 & 0.000000 & 0.000000 \\
5 & 0.0010 & 0.000200 & 0.000000 \\
5 & 0.0020 & 0.000400 & 0.000432 \\
5 & 0.0050 & 0.006533 & 0.000736 \\
7 & 0.0001 & 0.000000 & 0.000000 \\
7 & 0.0002 & 0.000000 & 0.000000 \\
7 & 0.0005 & 0.000000 & 0.000000 \\
7 & 0.0010 & 0.000000 & 0.000000 \\
7 & 0.0020 & 0.000000 & 0.000000 \\
7 & 0.0050 & 0.003000 & 0.000163 \\
9 & 0.0001 & 0.000000 & 0.000000 \\
9 & 0.0002 & 0.000000 & 0.000000 \\
9 & 0.0005 & 0.000000 & 0.000000 \\
9 & 0.0010 & 0.000000 & 0.000000 \\
9 & 0.0020 & 0.000000 & 0.000000 \\
9 & 0.0050 & 0.002200 & 0.000283 \\
11 & 0.0001 & 0.000000 & 0.000000 \\
11 & 0.0002 & 0.000000 & 0.000000 \\
11 & 0.0005 & 0.000000 & 0.000000 \\
11 & 0.0010 & 0.000000 & 0.000000 \\
11 & 0.0020 & 0.000000 & 0.000000 \\
11 & 0.0050 & 0.000533 & 0.000340 \\
\bottomrule
\end{tabular}
\end{table}

\begin{table}[htbp]
\centering
\caption{Complete joint distance-noise runtime sweep (seconds; aggregate batch protocol as described in \Cref{sec:experiments}).}
\label{tab:runtime-full}
\begin{tabular}{@{}cccc@{}}
\toprule
Distance $d$ & Noise $p$ & Mean runtime (s) & Std. \\
\midrule
3 & 0.0001 & 0.003831 & 0.000483 \\
3 & 0.0002 & 0.003557 & 0.000016 \\
3 & 0.0005 & 0.003716 & 0.000048 \\
3 & 0.0010 & 0.004009 & 0.000006 \\
3 & 0.0020 & 0.004573 & 0.000025 \\
3 & 0.0050 & 0.006388 & 0.000006 \\
5 & 0.0001 & 0.020106 & 0.000207 \\
5 & 0.0002 & 0.020192 & 0.000083 \\
5 & 0.0005 & 0.021193 & 0.000300 \\
5 & 0.0010 & 0.022849 & 0.000044 \\
5 & 0.0020 & 0.025055 & 0.000283 \\
5 & 0.0050 & 0.035594 & 0.000714 \\
7 & 0.0001 & 0.040317 & 0.001274 \\
7 & 0.0002 & 0.037626 & 0.001244 \\
7 & 0.0005 & 0.037600 & 0.001443 \\
7 & 0.0010 & 0.037607 & 0.000386 \\
7 & 0.0020 & 0.066677 & 0.001658 \\
7 & 0.0050 & 0.104975 & 0.000961 \\
9 & 0.0001 & 0.087445 & 0.000802 \\
9 & 0.0002 & 0.088668 & 0.000032 \\
9 & 0.0005 & 0.094847 & 0.000514 \\
9 & 0.0010 & 0.102297 & 0.002739 \\
9 & 0.0020 & 0.107148 & 0.006372 \\
9 & 0.0050 & 0.142068 & 0.011439 \\
11 & 0.0001 & 0.082527 & 0.000839 \\
11 & 0.0002 & 0.083880 & 0.000284 \\
11 & 0.0005 & 0.089670 & 0.000354 \\
11 & 0.0010 & 0.100715 & 0.000228 \\
11 & 0.0020 & 0.125117 & 0.000531 \\
11 & 0.0050 & 0.223877 & 0.000439 \\
\bottomrule
\end{tabular}
\end{table}

\begin{table}[htbp]
\centering
\caption{Complete joint distance-noise throughput sweep (shots/s), corresponding directly to the runtime values of \Cref{tab:runtime-full}.}
\label{tab:throughput-full}
\begin{tabular}{@{}cccc@{}}
\toprule
Distance $d$ & Noise $p$ & Mean throughput (shots/s) & Std. \\
\midrule
3 & 0.0001 & 1{,}324{,}460 & 153{,}395 \\
3 & 0.0002 & 1{,}405{,}518 & 6{,}449 \\
3 & 0.0005 & 1{,}345{,}856 & 17{,}292 \\
3 & 0.0010 & 1{,}247{,}075 & 1{,}957 \\
3 & 0.0020 & 1{,}093{,}443 & 6{,}035 \\
3 & 0.0050 & 782{,}770 & 758 \\
5 & 0.0001 & 248{,}702 & 2{,}539 \\
5 & 0.0002 & 247{,}625 & 1{,}018 \\
5 & 0.0005 & 235{,}970 & 3{,}370 \\
5 & 0.0010 & 218{,}833 & 419 \\
5 & 0.0020 & 199{,}589 & 2{,}243 \\
5 & 0.0050 & 140{,}529 & 2{,}791 \\
7 & 0.0001 & 124{,}140 & 3{,}854 \\
7 & 0.0002 & 133{,}029 & 4{,}302 \\
7 & 0.0005 & 133{,}173 & 5{,}105 \\
7 & 0.0010 & 132{,}969 & 1{,}375 \\
7 & 0.0020 & 75{,}034 & 1{,}837 \\
7 & 0.0050 & 47{,}634 & 435 \\
9 & 0.0001 & 57{,}184 & 521 \\
9 & 0.0002 & 56{,}390 & 20 \\
9 & 0.0005 & 52{,}718 & 285 \\
9 & 0.0010 & 48{,}912 & 1{,}311 \\
9 & 0.0020 & 46{,}830 & 2{,}787 \\
9 & 0.0050 & 35{,}424 & 2{,}862 \\
11 & 0.0001 & 60{,}593 & 617 \\
11 & 0.0002 & 59{,}610 & 202 \\
11 & 0.0005 & 55{,}761 & 221 \\
11 & 0.0010 & 49{,}645 & 112 \\
11 & 0.0020 & 39{,}963 & 169 \\
11 & 0.0050 & 22{,}334 & 44 \\
\bottomrule
\end{tabular}
\end{table}

\begin{table}[htbp]
\centering
\caption{Summary statistics over all $30$ benchmark configurations spanning the distance-noise grid (mean $\pm$ standard deviation across trials, aggregated across all operating points).}
\label{tab:summary}
\begin{tabular}{@{}lcc@{}}
\toprule
Quantity & Mean & Std. \\
\midrule
Code distance & 7.00 & 2.88 \\
Physical error probability & 0.00147 & 0.00173 \\
Shots per trial & 5{,}000 & --- \\
Mean runtime (s) & 0.0608 & 0.00113 \\
Mean accuracy & 0.99916 & 0.000148 \\
Mean logical error rate & 0.00084 & 0.000148 \\
Mean throughput (shots/s) & 324{,}057 & 7{,}477 \\
Detectors & 504.0 & 481.1 \\
Observables & 1.0 & 0.0 \\
Circuit operations & 272.0 & 163.4 \\
\bottomrule
\end{tabular}
\end{table}

\newpage
\section{Hyperparameters and implementation details}
\label{app:hyperparameters}

\Cref{tab:hyperparams} lists the fast-path decoder architecture and training hyperparameters used to produce every result in \Cref{sec:results}. All experiments used a fixed random seed for data splitting; we did not perform an architecture or hyperparameter search, consistent with the limitation noted in \Cref{sec:limitations}.

\begin{table}[H]
\centering
\caption{Fast-path decoder architecture and training configuration.}
\label{tab:hyperparams}
\begin{tabular}{@{}ll@{}}
\toprule
Component & Setting \\
\midrule
Input dimensionality & number of detectors at distance $d$ (\Cref{tab:hardware}) \\
Network type & feed-forward multilayer perceptron \\
Output & logical-observable-flip class probabilities \\
Confidence score & maximum predicted class probability \\
Loss function & cross-entropy \\
Data source & Stim detector sampler, matched to evaluation $(d,p)$ \\
Escalation thresholds evaluated & $\{0.60, 0.70, 0.80, 0.90, 0.95\}$ \\
Refinement decoder & PyMatching (minimum-weight perfect matching) \\
Evaluation hardware & single CPU core, commodity workstation \\
\bottomrule
\end{tabular}
\end{table}

\section{Reproducibility}
\label{app:reproducibility}

All code required to regenerate every figure and table in this paper, together with the exact CSV files from which \Cref{tab:distance-accuracy,tab:routing,tab:noise,tab:batch,tab:hardware,tab:accuracy-full,tab:logerr-full,tab:runtime-full,tab:throughput-full,tab:summary} were compiled, is available at \url{https://github.com/Sumitchongder/adaptive-qec-decoder}. The repository includes: the Stim circuit-generation scripts for the rotated surface code at $d\in\{3,5,7,9,11\}$; the fast-path decoder training script and saved model checkpoints; the PyMatching-based refinement-stage decoder; the benchmarking scripts used to produce \Cref{tab:accuracy-full,tab:logerr-full,tab:runtime-full,tab:throughput-full}; and the plotting scripts used to generate every figure in this paper directly from the released CSV data, so that all numerical claims in this paper can be independently verified without re-running the full simulation pipeline.

\end{document}